\pdfoutput=1
\documentclass[
  aps,pre,
  preprint,           
  nofootinbib,
  superscriptaddress,
  longbibliography
]{revtex4-2}
\usepackage[T1]{fontenc}
\usepackage[utf8]{inputenc}
\usepackage{microtype}
\usepackage{siunitx} 
\providecommand{\qty}[2]{\SI{#1}{#2}}
\providecommand{\ang}[1]{\SI{#1}{\degree}}
\usepackage{tabularx} 
\makeatletter
\@ifpackageloaded{inputenc}{%
  \DeclareUnicodeCharacter{2009}{\,}%
}{}
\makeatother
\usepackage[section]{placeins} 
\usepackage{mathtools}           
\usepackage{amsthm}              
\usepackage{newtxtext}
\usepackage[amsthm]{newtxmath}   
\usepackage{bm}
\mathtoolsset{showonlyrefs}

\newcommand{\Hb}{H_b}             
\usepackage{graphicx}
\graphicspath{{fig/}{figures/}}
\usepackage[most]{tcolorbox}
\usepackage{booktabs}
\usepackage{enumitem}
\DeclareSIUnit\bit{bit}
\DeclareSIUnit\nat{nat}
\usepackage{empheq}
\usepackage[caption=false]{subfig}

\usepackage{nameref}
\usepackage[colorlinks=true,
  linkcolor=blue,
  citecolor=blue,
  urlcolor=blue]{hyperref}

\theoremstyle{plain}

\theoremstyle{definition}

\begin{document}

\title{Murray’s Law as an Entropy-per-Information-Cost Extremum}

\author{Justin Bennett}
\email{jmb15@illinois.edu}
\affiliation{Department of Physics, University of Illinois Urbana--Champaign, Urbana, Illinois 61801, USA}
\thanks{Independent work; not conducted under, nor funded by, the University of Illinois research program.}

\date{\today}
\begin{abstract}
Transport networks must balance viscous pumping losses with the energetic cost of maintaining an operative architecture. This paper formulates that trade-off as an entropy–per–information–cost (EPIC) extremum that prices structural upkeep in calibrated units (joules per bit). An upkeep law \(r^{m}\) distinguishes volume-priced (\(m{=}2\)) from surface-priced (\(m{=}1\)) maintenance. In laminar Poiseuille flow, stationarity yields (i) a generalized Murray scaling \(Q\propto r^{\alpha}\) with \(\alpha=(m{+}4)/2\); (ii) a tariff‑weighted vector balance that fixes bifurcation geometry and predicts near‑symmetric daughter openings of \(\approx75^\circ\) for \(m{=}2\) and \(\approx97^\circ\) for \(m{=}1\); and (iii) a universal partition of power between pumping and upkeep. Eliminating radii gives a strictly concave flux cost \(Q^{\gamma}\) with \(\gamma=2m/(m{+}4)\), favoring mergers and deep tree hierarchies, and defines a routing index that induces Snell‑like refraction of optimal paths across spatial tariff contrasts. A preregistered, held‑out test on retinal bifurcations from the High–Resolution Fundus dataset (\(N=19{,}126\)) shows sharp vector closure: the median residual is \(R=0.232\) with a nonparametric 95\% bootstrap interval \([0.229,\,0.236]\), \(91\%\) of junctions fall under the pre‑specified strict threshold, and structure‑preserving nulls shift decisively to larger residuals. These results render classical branching relations explicitly unit‑bearing (J/bit) and provide falsifiable geometric targets and quantitative design rules for transport networks.
\end{abstract}

\maketitle

\newpage
\section{Introduction}
Nature channels free energy through architectural pathways that must be continually maintained, and the energetic burden of that upkeep is repaid by an increased \emph{entropy–production rate}—the volumetric rate of irreversible entropy production \cite{deGrootMazur,Onsager1931a,Onsager1931b,Seifert2012}. At laminar bifurcations this trade‑off leaves a geometric trace: Murray’s law, the empirically robust cubic relation linking one parent radius to its two daughters \cite{Murray1926a,Murray1926b,Sherman1981,Zamir1976Optimality}. Such systems balance viscous dissipation, which rises steeply as channels narrow, against the metabolic cost of building and perfusing larger conduits, selecting radii near a compromise where neither loss dominates. The present formulation recasts that compromise as a single extremum: the canonical cubic law, together with its surface‑weighted generalizations, emerges as the stationarity condition of an empirically calibratable \emph{entropy‑per‑information‑cost} (EPIC) variational principle. The resulting geometric and dynamical predictions are tested on retinal vascular junctions extracted from the High–Resolution Fundus (HRF) \cite{Budai2013IJBI}.

The analysis proceeds in the \emph{Poiseuille regime}, where steady, isothermal Newtonian flow isolates the trade‑off between viscous dissipation and the energetic cost of maintaining structure at a single branching site. In this near‑equilibrium setting—one parent feeding two daughters with fixed demands, \(Q_0=Q_1+Q_2\)—the Y‑junction serves as a controlled laboratory: infinitesimal changes in radii and branch angles map linearly, via the Hagen–Poiseuille law, to power expenditure, enabling sharp, falsifiable tests of EPIC predictions against empirical morphologies and fluxes.

This paper does \textbf{not} introduce a new law of transport; rather, it recasts the Murray optimum in a unit–bearing, information–thermodynamic ledger. In contrast to “maximum entropy production’’ heuristics, which posit that nonequilibrium systems maximize \emph{total} entropy production under coarse constraints \cite{MartyushevSeleznev2006PhysRep}, EPIC keeps the hydrodynamics fixed and prices the reliably maintained distinctions that keep a conduit operative in absolute energetic units via an \emph{effective bit tariff} \(\varepsilon_b=\zeta\,k_B T\ln 2\), anchored in Landauer’s bound and its experimental verification by Bérut \emph{et\,al.} \cite{Berut2012}. In these units the local dissipation \(\sigma_s T\) defines an information–throughput budget
\(\Psi_b \equiv \sigma_s T / \varepsilon_b\) \([\si{\bit\per\metre\cubed\per\second}]\), so the
transport law remains classical while the objective changes: geometry is selected by extremizing entropy production \emph{per unit information cost}, rather than by maximizing
\(\sigma_s T\) itself. When the tariff is spatially uniform and the structural bit budget is held fixed across admissible variations, the EPIC extremum collapses to the classical minimum–power design at fixed daughter flows, recovering the cubic form of Murray’s law. Definitions and units are summarized in \S\ref{sec:notation}.

Three main consequences follow from this reframing. \textbf{First}, the equal–marginal–price stationarity condition enforces a branchwise
“priced–flux’’ invariant \(Q_i^{2}/(c_i r_i^{m+4})=\mathrm{const}\), which implies the
flow–radius scaling \(Q \propto r^{\alpha}\) with \(\alpha=(m+4)/2\) and, by continuity at
the node, the tariff–weighted Murray relation
\(\sqrt{c_{0}}\,r_{0}^{\alpha}=\sqrt{c_{1}}\,r_{1}^{\alpha}+\sqrt{c_{2}}\,r_{2}^{\alpha}\). For homogeneous (volume–priced) tariffs \(c_0=c_1=c_2\) with \(m=2\), this reduces to the classical cubic law \(r_{0}^{3}=r_{1}^{3}+r_{2}^{3}\); for surface–priced upkeep \(m=1\), it yields the familiar subcubic analogue with \(\alpha=5/2\).

\textbf{Second}, translational neutrality implies a tariff-weighted vector closure
\(\sum_i c_i\,r_i^{m}\,\mathbf e_i=0\), in which each branch carries a ``bit-tension''
\(c_i r_i^{m}\mathbf e_i\). This balance fixes daughter opening angles: for given radii and upkeep exponent \(m\), the three weighted direction vectors must close, and the normalized misclosure provides a direct geometric diagnostic. Under homogeneous tariffs \(c_i=c\), this closure (up to an overall factor) reduces to the angle–area vector balance derived by Durand and co‑workers \cite{Durand2006PRE,Durand2007PRL}, yielding the characteristic \(\sim\!\ang{75}\) symmetric arterial bifurcation for \(m=2\). In what follows, the relation is applied at degree‑3 retinal junctions in HRF and compared against structure‑preserving null models; the closure itself is classical, while EPIC ties its weights to an audited informational price, permitting interpretation in absolute energetic units (J/bit).

\textbf{Third}, pricing dissipation at the informational tariff yields a two-faced “growth–cone’’ bound on the rate at which the stock of reliably decodable, stably maintained distinctions can increase,
\[
\dot I \;\le\; \min\!\Big\{\;\int_V \Psi_b\,\mathrm dV,\;\int_{\Sigma} \mathcal C_{\epsilon}\,\mathrm dA\Big\},
\]
where \(\int_V \Psi_b\,\mathrm dV\) is the interior bit budget purchasable from entropy production at the audited tariff and \(\int_{\Sigma} \mathcal C_{\epsilon}\,\mathrm dA\) is the boundary’s reliable information throughput at the stated blocklength and error probability. The operative constraint is explicit: at any instant, either the available power (via \(P_{\rm flow}/\varepsilon_b\)) or the finite communication capacity of the enclosing surface limits how rapidly new, reliably decodable degrees of freedom can be written into the domain.

These consequences of EPIC sit within, and are constrained by, a substantial literature on optimal transport networks. Minimizing global dissipation under surface/volume constraints yields (i) generalized Murray relations and (ii) a \emph{vectorial force–balance} angle law written as a closure of weighted branch vectors—the PRE/PRL benchmark of Durand and co-workers \cite{Durand2006PRE,Durand2007PRL}. At the network scale, related PRL work by Bohn \& Magnasco and by Katifori \emph{et\,al.} shows that \emph{concave} transport costs favor deep tree hierarchies, while damage or fluctuations generate loops, reproducing the observed tree–versus–loop phenomenology in optimal networks \cite{BohnMagnasco2007PRL,Katifori2010PRL}. The novelty here is not the geometric closure itself, but the explicitly unit–bearing \emph{EPIC ledger} (J/bit) that prices structure together with a domain–level \emph{growth–cone} bound on the rate of creating reliably decodable structure. Taken together, these additions yield unit–consistent power partitions, a concave effective edge cost with exponent \(\gamma=2m/(m{+}n)\) that explains the energetic preference for mergers when \(m<n\), and an \emph{EPIC index} \(\mathfrak n(x)\) defined in Eq.~\eqref{eq:EPIC_index_def}
(reducing to \([\varepsilon_b(x)c(x)/(\varepsilon_{b0}c_{0})]^{\,n/(m+n)}\) when \(\kappa_{\mathrm{hyd}}\) is spatially uniform) that induces Snell–type refraction of optimal routes across spatial tariff contrasts. Empirically, the junction–angle closure is evaluated on HRF using held–out directions and preregistered structure–preserving nulls, providing a falsifiable, image–level check that anchors these design–level predictions in measured retinal geometry.

The remainder of the paper proceeds as follows. The EPIC ledger section introduces the effective bit tariff and the throughput \(\Psi_b\). A \emph{Classical benchmark} (pumping
versus upkeep at fixed demand) fixes notation. The \emph{EPIC extremum} at a node is then derived, yielding the priced-flux invariant, the weighted Murray relation, and the translation-based angle closure. \emph{Empirical Tests} evaluate vector closure on HRF using preregistered nulls. The \emph{Growth–cone bound} section then states a domain-level rate limit on the creation of reliably decodable structure. \emph{Design corollaries (“EPIC optics”)} develop equipartition under dilation, the concave effective edge cost and its tree preference, and the Snell-type routing index in heterogeneous price fields. Unit-bearing definitions appear in \S\ref{sec:notation}.

\section{EPIC ledger: tariffs, structural load, and information throughput}
This section sets the EPIC ledger within the thermodynamics–of–information framework (for a review see Parrondo, Horowitz, and Sagawa \cite{Parrondo2015NatPhys}) by putting energy use and morphology on a common quantitative footing: both are expressed in units of reliably decoded bits. Here a “bit’’ is understood operationally as a single, reproducible binary decision (open/closed, high/low, etc.) resolved over a specified time window at a declared error tolerance. Three primitives carry out this change of variables. An \emph{information tariff} assigns an absolute energetic price to each such bit; a \emph{structural bit–rate} counts the rate at which a network must refresh, monitor, and stabilize its conduits; and a \emph{power–priced information throughput} re-expresses local dissipation as the rate at which reliable distinctions can, in principle, be maintained. Written in these units, energetic supply and informational demand live in the same currency, fixing the scale for the subsequent variational statements and enabling direct, testable links between geometry, power, and information.

The tariff is quantified by an effective bit energy
\[
\varepsilon_b=\zeta\,k_B T\ln 2,\qquad \eta_b\equiv \varepsilon_b^{-1},
\]
with \(T\) the (locally) isothermal audit temperature and \(\zeta\!\ge\!1\) a dimensionless overhead factor that aggregates finite–time, reliability, redundancy, and housekeeping costs, calibrated operationally in \S\ref{sec:calibration} (cf.\ Landauer’s bound \(k_B T\ln 2\) per reliable bit, experimentally verified by Bérut \emph{et\,al.} \cite{Berut2012}). Operationally, \(\varepsilon_b\) is the measured ratio of baseline-subtracted control-layer power to the stream of reliably decoded bits at the chosen error tolerance; its reciprocal \(\eta_b\) converts power density into an information-throughput density \(\Psi_b \equiv \sigma_s T/\varepsilon_b\) [\si{bit\,m^{-3}\,s^{-1}}], so that the familiar dissipation \(\sigma_s T\) is read as the rate of reliable information available to sustain structure.

On the demand side, the rate at which a branching conduit must maintain distinctions along a segment of radius \(r_i\) and length \(\ell_i\) is modeled as
\[
\mathcal B(\mathbf r)=\sum_{i=0}^{2} c_i\,\ell_i\,r_i^{m}, \qquad [\mathcal B]=\si{bit\,s^{-1}},
\]
where the positive coefficients \(c_i\) encode platform-specific sensing, control, and refresh burdens per unit surface or volume, and the exponent \(m\) captures how upkeep scales with size: \(m=2\) for volume-priced maintenance and \(m=1\) for surface-priced burdens. Pricing dissipation at the audited tariff defines a \emph{power–priced} information–throughput density,
\[
\Psi_b=\frac{\sigma_s T}{\varepsilon_b}\quad\bigl[\si{\bit\per\metre\cubed\per\second}\bigr],
\]
which can be interpreted as a continuous information–flow field in the stochastic–thermodynamics sense, where energetic and informational currents are explicitly coupled \cite{HorowitzEsposito2014PRX}. A brief note on thermodynamic uncertainty relations (TUR)—and why they are not used here without additional dynamical assumptions—appears in Appendix~\ref{app:TUR_remark}. 

Each watt of irreversible power is thereby converted into \(\eta_b=\varepsilon_b^{-1}\) reliably decodable bits per second. In a steady, isothermal control volume \(V\),
\[
\int_V \Psi_b\,\mathrm dV
=\frac{\int_V \sigma_s T\,\mathrm dV}{\varepsilon_b}
=\frac{P_{\rm flow}}{\varepsilon_b},
\]
so the hydraulic power \(P_{\rm flow}\) is read directly in \si{bit\,s^{-1}} at the audited price \(\varepsilon_b\). Because \(\sigma_s T\) is fixed by transport while \(\varepsilon_b\) is calibrated by an explicit control–layer audit, \(\Psi_b\) is both physically grounded and experimentally accessible. A numerical example that converts a measured pumping power into a bit–per–second budget, \(P_{\rm flow}/\varepsilon_b\), is given in \S\ref{sec:calibration}.

In this language, the EPIC ledger makes the design problem explicit: geometry is chosen to balance an audited \emph{supply} of reliable information,
\(\int_V \Psi_b\,\mathrm dV\), against the \emph{demand} imposed by structural upkeep, \(\mathcal B(\mathbf r)=\sum_i c_i \ell_i r_i^m\). The node extremum corresponds to extremizing entropy throughput \emph{per unit information cost} under this maintenance load, rather than minimizing dissipation alone. The audit protocol (choice of window \(\tau\)), the decomposition of \(\zeta\), and the reporting conventions that connect \(P_{\rm flow}\), \(\varepsilon_b\), and \(\Psi_b\) are detailed in \S\ref{sec:calibration}.

\section{Classical benchmark: pumping versus upkeep at fixed daughter demands}

Before deriving Murray–type relations within the EPIC framework, it is useful to recall the classical optimization that balances viscous pumping power against an explicit upkeep cost at \emph{fixed} daughter demands. Consider a single Y–junction in which one parent segment \((r_0,\ell_0)\) feeds two daughters \((r_1,\ell_1)\) and \((r_2,\ell_2)\). The fluid is taken incompressible and Newtonian; flow is steady, laminar, and fully developed in long circular tubes (entrance effects neglected). Under these conditions each segment obeys the Hagen–Poiseuille relations
\begin{equation}
\label{eq:poiseuille_classic}
Q_i=\frac{\pi r_i^{4}}{8\mu \ell_i}\,\Delta p_i,
\qquad
P_{{\rm flow},i}=Q_i\,\Delta p_i=\frac{8\mu \ell_i}{\pi}\,\frac{Q_i^{2}}{r_i^{4}},
\end{equation}
with dynamic viscosity \(\mu\) and volumetric flow \(Q_i\) (see, e.g., White and Bruus~\cite{WhiteViscous,BruusMicrofluidics}). Throughout this benchmark, daughter demands are prescribed so that \(Q_0=Q_1+Q_2\) and segments are isothermal. As a contrasting ensemble, note that if \(\Delta p\) were fixed at the node, no nontrivial optimum in the radii would exist, since \(P_{\rm flow}\propto r^{4}\) would grow monotonically with size~\cite{deGrootMazur}.

To represent the standing energetic burden of keeping a patent conduit and its working fluid available, introduce an upkeep power that scales as a power of radius,
\begin{equation}
\label{eq:upkeep_classic}
P_{{\rm maint},i}=\Lambda_m\,\ell_i\,r_i^{m},
\qquad
m=
\begin{cases}
2,&\text{volume–priced},\\
1,&\text{surface–priced},
\end{cases}
\end{equation}
with \(\Lambda_m>0\). At fixed \(Q_i\) the per–segment objective separates,
\begin{equation}
\label{eq:Pi_sum}
P_i(r_i)=\frac{8\mu \ell_i}{\pi}\frac{Q_i^{2}}{r_i^{4}}
+\Lambda_m\,\ell_i\,r_i^{m},
\end{equation}
so each radius can be optimized independently of the others given its prescribed flow. In the classical benchmark, \(\Lambda_m\) is an \emph{ad hoc} scale that stands in for the price of keeping structure. In EPIC the priced upkeep power is denoted \(P_{\rm struct}\), with
\(P_{{\rm struct},i}\equiv \Lambda\,c_i\,\ell_i\,r_i^{m}\), so the classical \(P_{{\rm maint},i}\) plays the same role. The EPIC formulation makes this term \emph{physical} by replacing \(\Lambda_m\,\ell_i r_i^{m}\) with an \emph{audited} cost \(\varepsilon_b\,c_i\,\ell_i r_i^{m}\), where \(\varepsilon_b\) (J/bit) is measured from control–layer power and reliably
decoded bit–rate, and \(c_i r_i^{m}\) counts the \emph{structural bit–rate} demanded by geometry, as overviewed in the next section.

Stationarity of \eqref{eq:Pi_sum} at fixed \(Q_i\) gives
\[
\frac{dP_i}{dr_i}
=-\frac{32\mu \ell_i}{\pi}\,\frac{Q_i^2}{r_i^{5}}
+m\,\Lambda_m\,\ell_i\,r_i^{m-1}=0,
\]
hence the branchwise invariant
\begin{equation}
\frac{Q_i^2}{r_i^{m+4}}=\frac{\pi m \Lambda_m}{32\mu}
\quad\Longrightarrow\quad
\label{eq:classic_alpha}
Q_i\propto r_i^{\alpha}\,,\qquad \alpha=\frac{m+4}{2}\,.
\end{equation}
Imposing continuity \(Q_0=Q_1+Q_2\) then yields the generalized Murray family
\begin{equation}
\label{eq:classic_Murray}
r_0^{\alpha}=r_1^{\alpha}+r_2^{\alpha}\,,\qquad \alpha=\tfrac{m+4}{2}\,.
\end{equation}
Thus the classical cubic law \(r_0^3=r_1^3+r_2^3\) arises when upkeep is volume–priced \((m=2)\), whereas a surface–priced burden \((m=1)\) produces the subcubic exponent \(\alpha=5/2\)~\cite{Murray1926a,Murray1926b,Sherman1981,Zamir1976Optimality,Emerson2006,BejanLorente2004}. The stationary point of \(P_i(r_i)\) is in fact a strict minimum. Differentiating \eqref{eq:Pi_sum} twice yields
\[
\frac{d^2P_i}{dr_i^2}
= \frac{160\,\mu\,\ell_i}{\pi}\,\frac{Q_i^2}{r_i^{6}}
+ m(m{-}1)\,\Lambda_m\,\ell_i\,r_i^{\,m-2},
\]
which is strictly positive for \(r_i>0\) and \(m\in\{1,2\}\). The optimizer in \eqref{eq:classic_alpha}–\eqref{eq:classic_Murray} is therefore unique. Moreover, multiplying the first–order condition by \(r_i\) exposes a transparent power partition,
\[
4\,P_{{\rm flow},i}=m\,P_{{\rm maint},i}
\quad\Longrightarrow\quad
P_{{\rm flow},i}=\frac{m}{4}\,P_{{\rm maint},i},
\]
so that at the optimum \(P_{\rm flow}=\tfrac12 P_{\rm maint}\) for \(m=2\) (volume–priced upkeep) and \(P_{\rm flow}=\tfrac14 P_{\rm maint}\) for \(m=1\) (surface–priced upkeep). Because \(\ell_i\) cancels from the stationarity condition, the scaling \eqref{eq:classic_alpha} and the additive law \eqref{eq:classic_Murray} continue to hold even when the segment lengths differ.

The validity domain of this benchmark is that of \eqref{eq:poiseuille_classic}: steady, laminar flow of a Newtonian fluid in long circular tubes. Outside this regime the same variational structure can be re–run with an appropriate dissipation law; for example, in small vessels the apparent viscosity becomes radius–dependent (Fåhræus–Lindqvist effect), \(\mu\to\mu(r)\), which shifts the effective exponent while preserving the underlying optimization logic~\cite{Pries1992,Secomb2013,Ascolese2019}. In this form the classical \emph{pumping \(+\) upkeep} calculation fixes the essentials of the node problem used below: an ensemble with fixed daughter demands \((Q_1,Q_2)\) under isothermal conditions, a separable per–segment objective at fixed flow, a unique interior optimum with \(Q\propto r^{(m+4)/2}\), and the additive law \(r_0^{\alpha}=r_1^{\alpha}+r_2^{\alpha}\). The only \emph{modeling} ingredient is the upkeep scale \(\Lambda_m\) in the term \(\Lambda_m\,\ell_i r_i^{m}\). EPIC replaces this ad hoc factor by an \emph{audited} pair \((\varepsilon_b,c_i)\) and a priced objective \(P_{\rm flow}+\Lambda\,\mathcal B\), with \(\mathcal B=\sum_i c_i\,\ell_i r_i^{m}\) and \(\Lambda=\lambda\,\varepsilon_b\), so that upkeep is expressed in calibrated units (J/bit) and counted explicitly as a structural bit–rate. When \(\varepsilon_b\) is uniform and \(\mathcal B\) is held fixed across admissible variations, the EPIC extremum reduces to the same minimum–power design at fixed daughter flows.

\section{EPIC extremum at a node}
Building on the classical benchmark, the EPIC extremum changes only the \emph{accounting}: transport dissipation and structural upkeep are expressed in a single audited currency, with entropy production priced at an effective bit energy that funds the reliably maintained distinctions keeping channels operative. Specializing this domain–level ledger to a steady Y–junction with fixed daughter demands yields a node objective whose Euler–KKT conditions provide the structure used throughout: equal–marginal–price stationarity reproduces \(Q\propto r^{(m+4)/2}\) and the tariff–weighted Murray relation; translating the junction produces a tariff–weighted vector closure that fixes the branch angles; and the same calculus delivers per–segment power partitions, coordinate–free diagnostics, and rheological generalizations. The subsections below proceed as follows: (i) domain-level variational program to node objective, (ii) equal–marginal–price condition, (iii) branch invariant and tariff–weighted Murray law, and (iv) angle selection by junction translation, with generalizations collected at the end of the section.

\subsection{Domain program to node objective}

At the level of a space–time domain $\mathcal D$, the design question is to choose a geometry that trades interior dissipation against the energetic price of maintaining reliable structure—expressed in a single, audited currency. This trade is written as an energy functional,
\begin{equation}
\label{eq:EPIC_variational}
\delta\!\bigg[
  \int_{\mathcal D}\sigma_s\,T\,dV\,dt
  \;+\;
  \lambda \int_{\mathcal D}\varepsilon_b\,\beta\,dV\,dt
\bigg] = 0,\qquad \lambda\ge 0,
\end{equation}
where $\sigma_s$ is the local entropy–production density, $T$ the node–isothermal audit temperature, and $\beta$ the density of reliably executed bit–operations. Both integrals carry units of energy: the first is the energy dissipated by transport within $\mathcal D$, the second is the energy \emph{spent} to sustain reliably decodable structure at the audited price $\varepsilon_b$ (J/bit). The multiplier $\lambda$ is therefore dimensionless: it regulates the exchange rate between “watts of dissipation’’ and “watts dedicated to reliable bits’’ in the extremum.

For a single, steady, isothermal node with uniform \(\varepsilon_b\), the spacetime ledger factorizes over an audit window of duration \(\tau\): \(\int_{\mathcal D}\sigma_s T\,dV\,dt=\tau\,P_{\rm flow}\) and \(\int_{\mathcal D}\varepsilon_b\,\beta\,dV\,dt=\tau\,\varepsilon_b\,\mathcal B\), where \(P_{\rm flow}\) is the hydraulic (dissipated) power and \(\mathcal B=\sum_{i=0}^{2}c_i\,\ell_i\,r_i^{m}\) is the structural bit–rate demanded by the geometry; dividing by \(\tau\) and absorbing the tariff into \(\Lambda\equiv\lambda\,\varepsilon_b>0\) yields the node-level objective
\begin{equation}
\label{eq:EPIC_additive_node}
\mathcal L(\mathbf r)=P_{\rm flow}(\mathbf r)+\Lambda\,\mathcal B(\mathbf r),
\end{equation}
posed at fixed daughter demands \(Q_1,Q_2\) with \(Q_0=Q_1+Q_2\). In this steady setting both terms carry units of power, so the geometry \(\mathbf r\) is chosen to balance a priced supply of reliable information (\(\Lambda\,\mathcal B\)) against the hydraulic cost (\(P_{\rm flow}\)); an equivalent isoperimetric (ratio) program that maximizes entropy throughput \emph{per} information cost is mathematically identical and leads to the same Euler–KKT conditions. Stationarity of \eqref{eq:EPIC_additive_node} with respect to each radius then enforces
\begin{equation}
\label{eq:stationarity_branch}
\frac{\partial P_{\rm flow}}{\partial r_i}+\Lambda\,\frac{\partial \mathcal B}{\partial r_i}=0
\qquad (i=0,1,2),
\end{equation}
which states that, at the optimizer, the marginal reduction in pumping power from an infinitesimal change in \(r_i\) is exactly offset by the tariff-priced marginal increase in structural load—the equal-marginal-price rule.

\subsection{Branch invariant and weighted Murray (Poiseuille)}

In the Poiseuille regime each branch carries a per–segment hydraulic cost \(P_{{\rm flow},i}=(8\mu \ell_i/\pi)\,Q_i^2 r_i^{-4}\) and a priced structural load \(\mathcal B_i=c_i\,\ell_i\,r_i^{m}\). Substituting these into the stationarity condition \eqref{eq:stationarity_branch} and cancelling the common length \(\ell_i>0\) yields an equal–marginal–price balance that condenses to the branchwise invariant
\begin{equation}
\label{eq:node_invariant}
\frac{Q_i^2}{c_i\,r_i^{m+4}}=\mathcal{K},
\qquad
\mathcal{K}\equiv\frac{\pi m\Lambda}{32\,\mu}>0,
\end{equation}
which is independent of segment length and has units \([\mathcal{K}]=\si{\metre^{3}\per\second\per\bit}\). Solving \eqref{eq:node_invariant} gives the flow–radius scaling
\begin{equation}
\label{eq:alpha}
Q_i=\mathcal{K}^{1/2}\,\sqrt{c_i}\;r_i^{\alpha},
\qquad
\alpha=\frac{m+4}{2},
\end{equation}
and enforcing continuity \(Q_0=Q_1+Q_2\) at the node produces the tariff–weighted Murray relation
\[
\sqrt{c_0}\,r_0^{\alpha}=\sqrt{c_1}\,r_1^{\alpha}+\sqrt{c_2}\,r_2^{\alpha}.
\]
This is the PRE benchmark: Durand derives a generalized Murray relation from a global minimum–dissipation principle under surface/volume constraints \cite{Durand2006PRE}.
Under homogeneous tariffs \(c_0=c_1=c_2\) this collapses to the familiar additive family \(r_0^{\alpha}=r_1^{\alpha}+r_2^{\alpha}\); in particular, volume–priced upkeep (\(m=2\)) gives \(\alpha=3\) (the classical cubic law) and surface–priced upkeep (\(m=1\)) gives \(\alpha=5/2\).

The same Euler condition exposes a transparent power partition. Multiplying \eqref{eq:stationarity_branch} by \(r_i\) and using the definitions above,
\[
4\,P_{{\rm flow},i}=m\,\Lambda\,c_i\,\ell_i\,r_i^{m}\equiv m\,P_{{\rm struct},i},
\]
so each branch satisfies \(P_{{\rm flow},i}=(m/4)\,P_{{\rm struct},i}\). Summing over branches gives the global split \(P_{\rm flow}=\tfrac12 P_{\rm struct}\) for \(m=2\) and \(P_{\rm flow}=\tfrac14 P_{\rm struct}\) for \(m=1\) (with \(P_{\rm struct}\equiv\Lambda\,\mathcal B\)), completing the link between the invariant, the Murray family, and the energetic balance at optimality.

\subsection{Angle selection by junction translation}
Let \(\mathbf e_i\) be unit vectors pointing outward along the three branches. An infinitesimal translation of the junction by \(\delta\mathbf x\) changes the segment lengths as \(\delta \ell_i=-\,\mathbf e_i\!\cdot\!\delta\mathbf x\). Stationarity of \(\mathcal L\) with respect to this translation requires \(\sum_i (\partial \mathcal L/\partial \ell_i)\,\mathbf e_i=\mathbf 0\). Using \(\partial \mathcal L/\partial \ell_i=(8\mu/\pi)\,Q_i^2 r_i^{-4}+\Lambda\,c_i r_i^{m}\) and eliminating the hydraulic prefactor with the radius stationarity \eqref{eq:stationarity_branch} collapses the balance to a \emph{tariff-weighted vector closure}
\begin{equation}
\label{eq:angle_vector}
\sum_{i=0}^2 c_i\,r_i^{m}\,\mathbf e_i=\mathbf 0.
\end{equation}
Angle selection appears as a vectorial force balance with weights tied to cross–sectional measures, equivalent to translating the junction (Durand’s Eq.~(6)) \cite{Durand2006PRE}. 
This is the EPIC analogue of the classical angle laws obtained from minimum-work arguments in vascular branching (cf.\ Zamir’s optimality analysis \cite{Zamir1976Optimality}) and is formally identical to force balance at a three-junction (Plateau–Young construction \cite{WeaireHutzler1999}), with with tariff weights \(a_i\equiv c_i r_i^{m}\).

Taking magnitudes in \eqref{eq:angle_vector} shows that the three “bit–tensions’’ \(a_i \equiv c_i r_i^{m}\) (for \(i=0,1,2\)) close a Euclidean triangle. The daughter opening angle \(\theta_{12}\) therefore obeys the usual cosine relation
\begin{equation}
\label{eq:angle_cosine}
\cos\theta_{12}
=\frac{a_0^{2}-a_1^{2}-a_2^{2}}{2\,a_1 a_2}\,.
\end{equation}
For homogeneous tariffs \(c_0=c_1=c_2\) with nearly symmetric daughters \(r_1=r_2\equiv r\), continuity together with \eqref{eq:alpha} gives \(r_0=2^{1/\alpha}r\). In this symmetric limit the vector closure reduces to \(a_0 = 2 a_1 \cos(\theta_{12}/2)\), so that
\begin{align}
\cos\!\Big(\tfrac{\theta_{12}}{2}\Big)
&= \frac{a_0}{2a_1}
 = \frac{r_0^{m}}{2 r^{m}}
 = \frac{(2^{1/\alpha} r)^{m}}{2 r^{m}} \\
&= 2^{\,\frac{m}{\alpha}-1}
 = 2^{\,\frac{2m}{m+4}-1}
 = 2^{\,\frac{m-4}{m+4}}.
\end{align}
Thus the parameter–free prediction can be written compactly as
\begin{equation}
\label{eq:angle_symmetric}
\cos\!\Big(\tfrac{\theta_{12}}{2}\Big)
=2^{\,\frac{2m}{m+4}-1}
=2^{\,\frac{m-4}{m+4}}\,,
\end{equation}
which yields \(\theta_{12}\!\approx\!\ang{74.93}\) for \(m{=}2\) (volume-priced) and \(\theta_{12}\!\approx\!\ang{97.44}\) for \(m{=}1\) (surface-priced). Notably, for a \emph{single} upkeep exponent \(m\), the closure \eqref{eq:angle_vector} depends only on \((m,c_i)\) and is \emph{independent} of the hydraulic exponent \(n\) in a generalized dissipation law \(P_{\rm flow}\propto Q^2 r^{-n}\); angles are then fixed entirely by structural pricing, not by rheology. When upkeep blends surface and volume contributions (per–length bit–rate \(a r^2+b r\)), the translation closure acquires the weights \(1+\tfrac{2}{n}\) (for \(r^2\)) and \(1+\tfrac{1}{n}\) (for \(r\)); for Poiseuille (\(n=4\)) this reduces to \(3/2\) and \(5/4\).

When local structural pricing blends surface and volume, with per–length bit–rate \(\mathcal B_i/\ell_i = a\,r_i^2 + b\,r_i\), the same translation step yields a modified, testable closure,
\begin{equation}
\label{eq:angle_mixed}
\sum_{i=0}^2 \Bigl(\bigl(1+\tfrac{2}{n}\bigr)a\,r_i^{2}+\bigl(1+\tfrac{1}{n}\bigr)b\,r_i\Bigr)\,\mathbf e_i=\mathbf 0,
\end{equation}
so that the junction angles are controlled by a specific mixed surface–volume weighting. This provides a characteristic mixed–pricing signature for angle selection in platforms where upkeep has both bulk (volume–like) and boundary (surface–like) contributions.

\subsection{Generalizations and rheology}
Two extensions clarify that junction angles are controlled purely by geometry and tariffs through the pair \((m,c_i)\), while rheology enters only through the flow–radius slope and shear–based invariants. First, allow a generalized per–length dissipation \(P_{\rm flow}\propto Q^{2} r^{-n}\) with \(n>0\) and structural pricing \(\mathcal B\propto r^{m}\) with \(m>0\). The node objective then separates into a transport term homogeneous of degree \(-n\) in \(r_i\) and a structural term homogeneous of degree \(m\). An infinitesimal translation of the junction changes each segment length as \(\delta \ell_i=-\mathbf e_i\!\cdot\!\delta\mathbf x\); taking the directional derivative along \(\delta\mathbf x\) and imposing stationarity removes all material prefactors and the hydraulic exponent \(n\), leaving the same tariff–weighted vector closure
\begin{equation}
\label{eq:angle_vector_repeat}
\sum_{i} c_i r_i^{m}\,\mathbf e_i=0,
\end{equation}
identical to the Poiseuille case. Stationarity with respect to each radius at fixed \(Q_i\) then relates the homogeneous degrees and yields the generalized flow–radius law \(Q \propto \sqrt{c}\,r^{(m+n)/2}\), i.e.\ \(\alpha=(m+n)/2\), so rheology (\(n\)) enters only through the slope \(\alpha\), not through the translation–derived closure.

A radius–dependent apparent viscosity \(\mu=\mu(r)\) likewise leaves the closure \eqref{eq:angle_vector_repeat} unchanged and modifies only the branch invariant and the effective flow–radius slope. At fixed \(Q_i\), stationarity yields the exact branchwise invariant
\begin{equation}
\label{eq:EPIC_invariant_mu_repeat}
\frac{Q_i^2}{c_i\,r_i^{m+4}}
=\frac{\pi m \Lambda}{8\,[\,4\,\mu(r_i)-r_i\,\mu'(r_i)\,]}
\equiv \mathcal{K}_{\rm eff}(r_i),
\end{equation}
which reduces to the Poiseuille result when \(\mu'\equiv 0\). Writing \(D(r)\equiv 4\mu(r)-r\mu'(r)\), a logarithmic derivative gives the exact local slope
\begin{equation}
\label{eq:alpha_eff_exact_repeat}
\alpha_{\rm eff}(r)
\equiv \frac{d\ln Q}{d\ln r}
=\frac{1}{2}\Big[(m{+}4)-\frac{d}{d\ln r}\ln D(r)\Big],
\end{equation}
so that if \(D(r)\) increases with \(r\) (i.e., \(\tfrac{d}{d\ln r}\ln D>0\)) then \(\alpha_{\rm eff}<\tfrac{m+4}{2}\), whereas if \(D(r)\) decreases with \(r\) then \(\alpha_{\rm eff}>\tfrac{m+4}{2}\). In particular, a monotonic increase of \(\mu(r)\) alone does not determine the sign: the shift is governed by \(D(r)=4\mu(r)-r\mu'(r)\). In the slow‑variation limit \(D(r)=4\mu(r)[1-\xi(r)]\) with \(|\xi|\ll 1\) and slowly varying \(\xi(r)\),
\begin{equation}
\label{eq:alpha_eff_slowmu_repeat}
\alpha_{\rm eff}(r)
=\frac{m+4}{2}-\frac{1}{2}\frac{d\ln\mu}{d\ln r}
+\mathcal O\!\Big(\xi,\ \xi^2,\ \tfrac{d\xi}{d\ln r}\Big),
\end{equation}
which recovers the expected trend when \(\mu(r)\) varies slowly~\cite{Pries1992,Secomb2013,Ascolese2019}.

The same rheological dependence can be recast as a second, shear–based invariant that is independent of branch length. Using \(\tau_w=4\,\mu(r)\,Q/(\pi r^{3})\) together with \eqref{eq:EPIC_invariant_mu_repeat} gives
\begin{equation}
\label{eq:priced_shear_invariant_repeat}
\mathscr S_i
\ \equiv\
\frac{\tau_{w,i}}{\sqrt{c_i}}\,
\frac{\sqrt{\,4\,\mu(r_i)-r_i\,\mu'(r_i)\,}}{\mu(r_i)}\,
r_i^{\frac{2-m}{2}}
\ =\ \sqrt{\frac{2 m \Lambda}{\pi}}\!,
\end{equation}
which takes the same value for \(i=0,1,2\). Thus \(\mathscr S_i\) packages geometry, tariffs, and rheology into a single branch–independent scalar: for equal tariffs \(c_i=c\) and constant viscosity \(\mu\), it reduces to constant wall shear when \(m=2\), and to constant \(\tau_w r^{1/2}\) when \(m=1\). The first case recovers the classical “uniform shear’’ optimality condition often invoked in vascular design \cite{Zamir1976Optimality}, while the second yields a distinct prediction appropriate to surface–priced upkeep. In practice, \(\mathscr S_i\) serves as a length–free stationarity diagnostic complementary to radius–flow scaling: systematic departures from a common \(\mathscr S_i\) across branches indicate that the junction is not at the EPIC optimum, that the assumed tariffs \(c_i\) or upkeep exponent \(m\) are mis–specified, or that the adopted rheological law \(P_{\rm flow}\propto Q^2 r^{-n}\) with \(\mu(r)\) is inadequate for the regime under study.

Finally, the Euler condition admits a coordinate-free diagnostic that does not require \(\Lambda\):
\begin{equation}
\label{eq:EPIC_segment_invariant_repeat}
\mathcal I_i\ \equiv\ \frac{\partial P_{\rm flow}/\partial r_i}{\partial \mathcal B/\partial r_i}=-\,\Lambda\quad (i=0,1,2),
\end{equation}
equivalently the requirement that \(Q_i^2/(c_i r_i^{m+4})\) be equal across branches (or \(Q_i^2/(c_i r_i^{m+4})\to\mathcal{K}_{\rm eff}(r_i)\) in the \(\mu(r)\) case via \eqref{eq:EPIC_invariant_mu_repeat}). When the informational tariff is uniform and admissible variations hold \(\mathcal B\) fixed, the EPIC extremum collapses to the classical minimum-pumping benchmark at fixed \((Q_1,Q_2)\), so that plug-flow, slip, shear-thinning, and \(\mu(r)\) variants reweight the flow–radius slope and priced-shear invariants but leave the translation-derived closure \eqref{eq:angle_vector_repeat} intact. This separation mirrors Durand’s PRE analysis, in which the relation between angles and cross-sectional areas appears as a vectorial force balance tied to geometric constraints rather than to the specific Poiseuille exponent~\cite{Durand2006PRE}. Full derivations, including the exact invariant for \(\mu(r)\), the local slope \(\alpha_{\rm eff}(r)\), and the priced–shear diagnostic \(\mathscr S_i\), are collected in Appendix~\ref{app:rheology}.

\section{Empirical tests}
To obtain a first, controlled test of EPIC’s geometric predictions, it is natural to work in a regime where its content sharp and unambiguous: individual vascular bifurcations. The aim is not to settle the full range of validity of the framework, but to probe one concrete consequence in a setting where the underlying fluid mechanics are well understood. Retinal degree-3 junctions offer a particularly clean laboratory: in the laminar Poiseuille regime they realize the classical minimum-work problem of a single parent vessel feeding two daughters, the configuration that underlies the analyses of Murray, Sherman, and Zamir~\cite{Murray1926a,Murray1926b,Sherman1981,Zamir1976}. At this scale, small perturbations of radii and opening angles map in a controlled way onto changes in viscous dissipation and structural upkeep, so the EPIC ledger reduces to a concrete node–level statement: for a given upkeep exponent \(m\), the three tariff–weighted flux vectors \(r_i^{m}\mathbf e_i\) at a Y–junction should close, with finite branch–specific tariffs \((c_0,c_1,c_2)\). The analysis below turns this into a quantitative, per-junction compatibility test: do real bifurcations behave as if their radii and directions were chosen to satisfy that balance when \(m\) is inferred solely from the scalar geometry \((r_0,r_1,r_2,\theta_{12})\) with directions held out, and is the resulting vector closure stable across images that differ in resolution, segmentation quality, and local branching morphology? The following seeks to address these questions using node-level closure residuals, comparisons to fixed-\(m\) and radii-only baselines, structure-preserving geometric nulls, and systematic quality-control and segmentation scans.

\subsection{Analysis methods}
\begin{figure*}[t]
  \centering
  \includegraphics[width=0.32\textwidth]{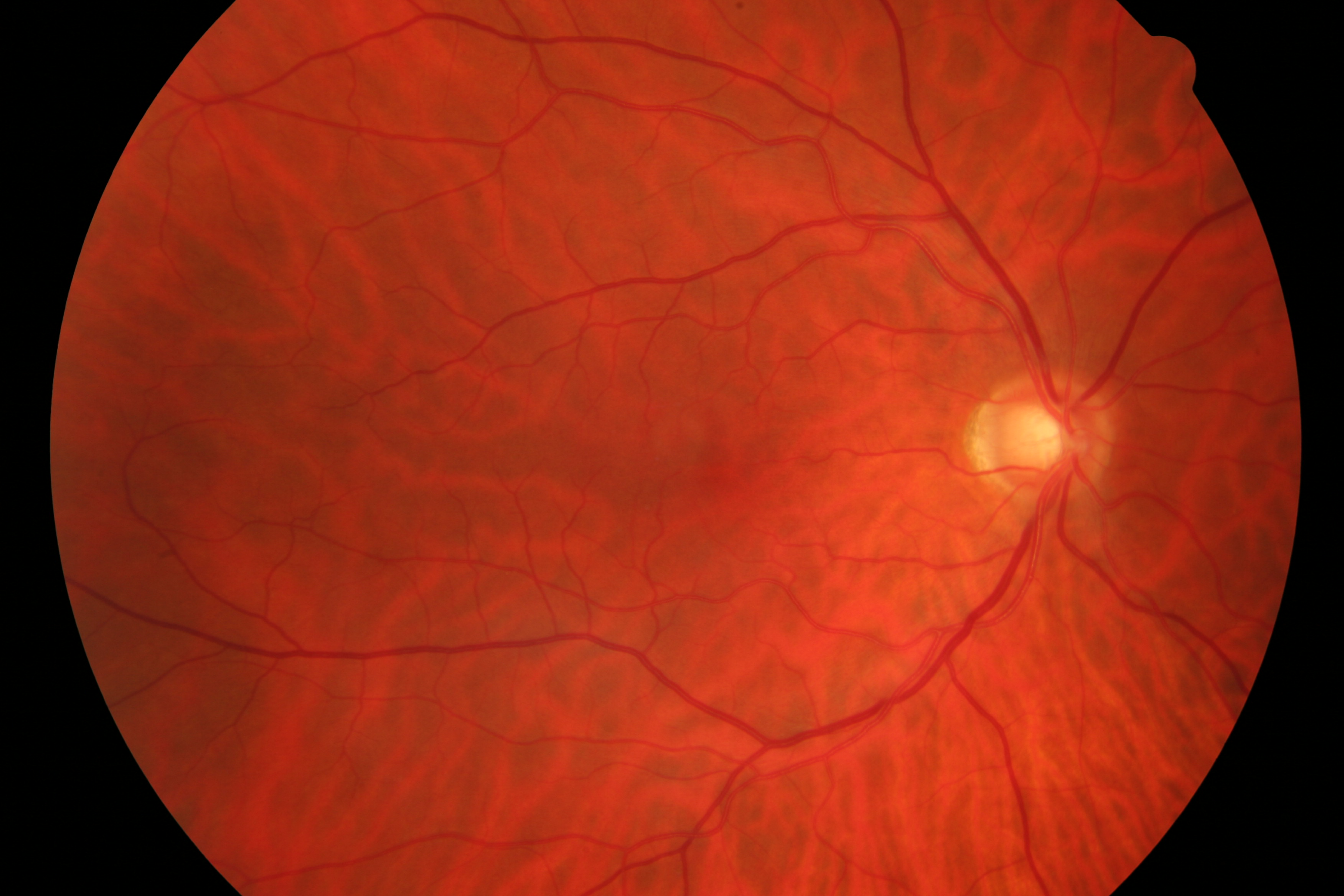}\hfill
  \includegraphics[width=0.32\textwidth]{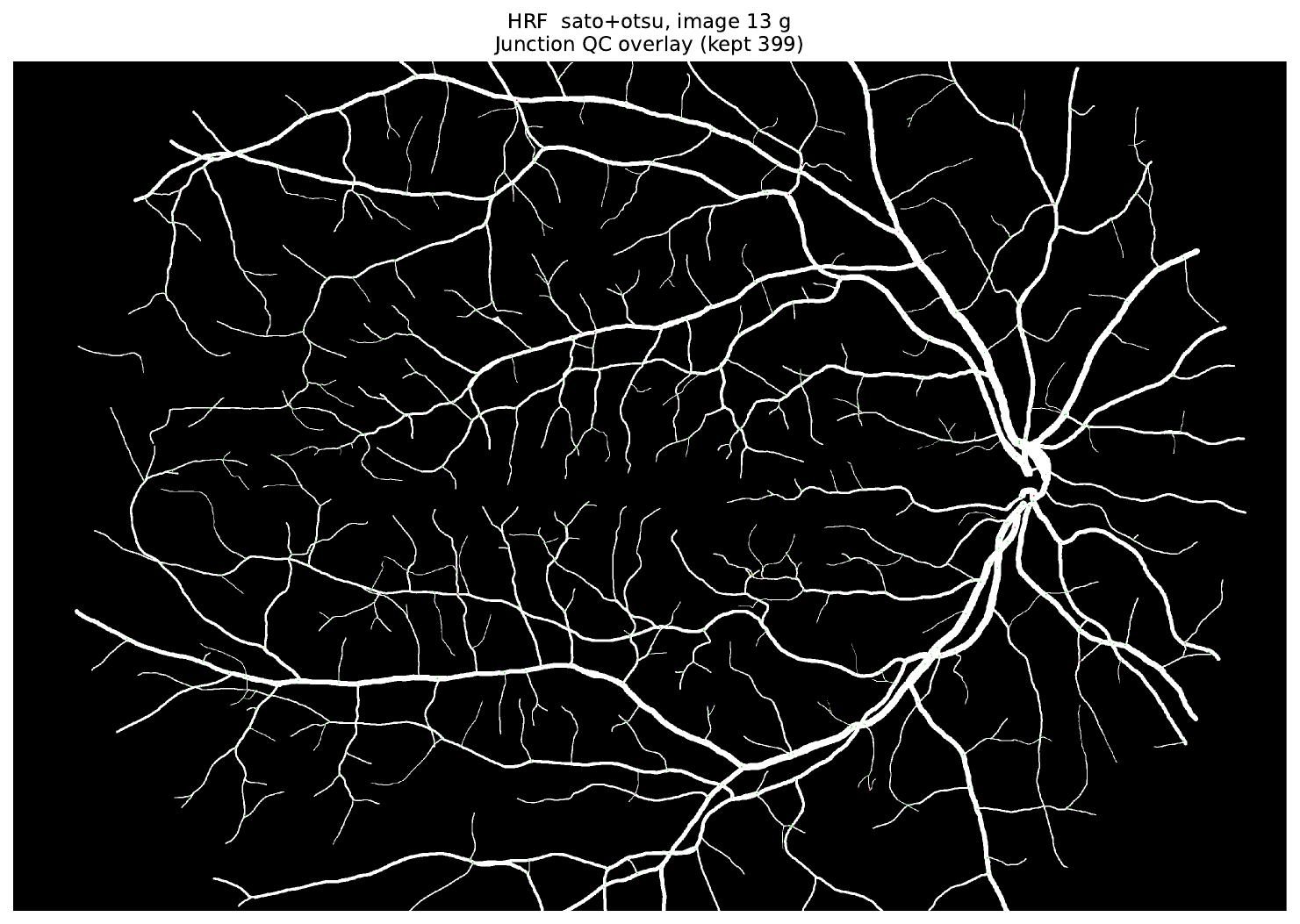}\hfill
  \includegraphics[width=0.32\textwidth]{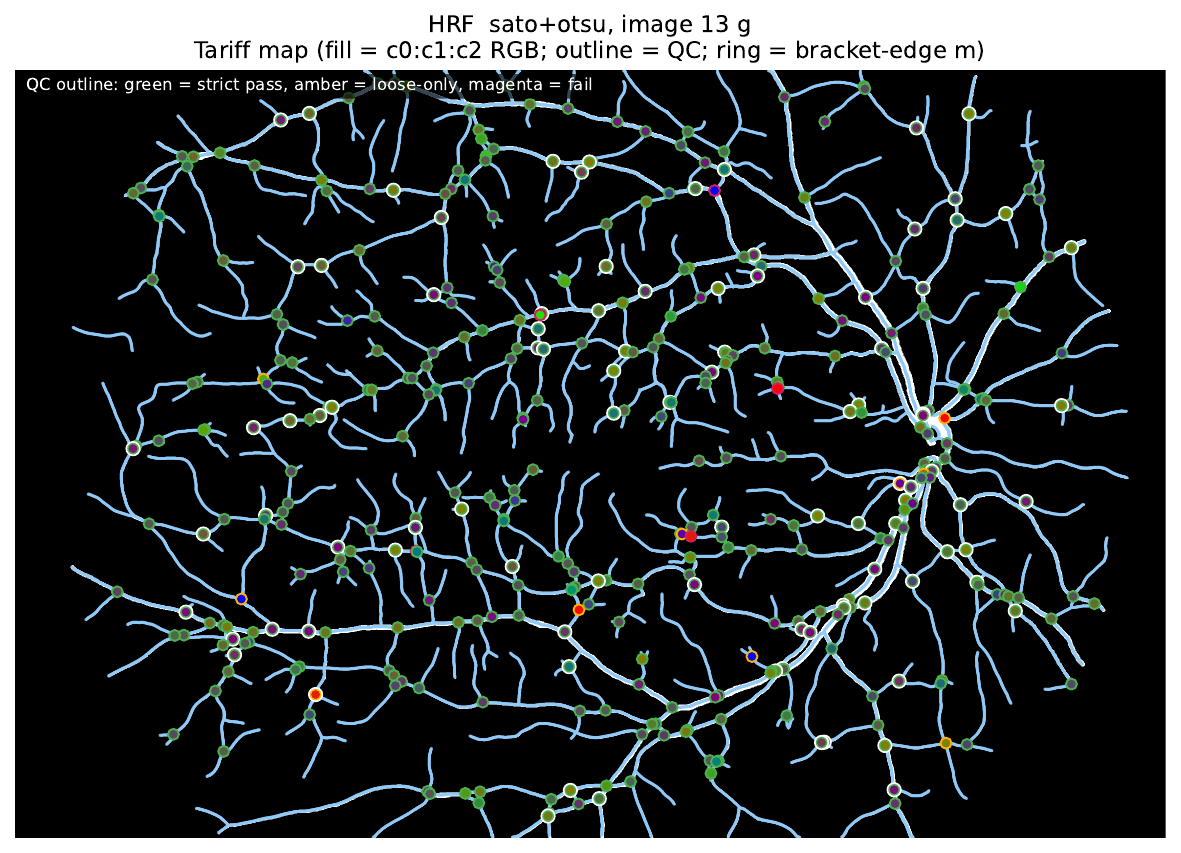}
  \caption{\textbf{From image to analyzable geometry (HRF).}
  (a) Original color fundus image (HRF, file \texttt{13\_g}). 
  (b) Binary vessel map from the normalized green channel after multiscale vesselness filtering, Otsu thresholding, and light morphological cleanup. 
  (c) Skeletonized vessel graph with per–junction tariffs and \emph{R‑band (interpretive) classes} overlaid. All images are drawn from the High--Resolution Fundus (HRF) Image Database (Pattern Recognition Lab, FAU), distributed under a Creative Commons Attribution 4.0 International (CC BY~4.0) license~\cite{HRF_web}; see Refs.~\cite{Budai2013IJBI,Odstrcilik2013IETIP} for dataset and label details.}
  \label{fig:hrf-open}
\end{figure*}

The High--Resolution Fundus (HRF) Image Database comprises 45 RGB fundus photographs (15 healthy, 15 with diabetic retinopathy, 15 with glaucoma). The corpus is publicly available under the Creative Commons Attribution 4.0 International (CC BY~4.0) license and is widely used in retinal vessel benchmarks~\cite{Budai2013IJBI,Odstrcilik2013IETIP}. Each image is reduced to a single–channel intensity field by extracting the green plane and applying a 1--99\,\% contrast normalization. Vessels are segmented from the normalized green channel using a multiscale \emph{Sato} line filter over $\sigma\in[1,8]$\,px, followed by \emph{Otsu} thresholding and light morphological cleanup to remove small components and holes. This \emph{Sato+Otsu} configuration is used for all HRF results reported here. The resulting binary mask is skeletonized by thinning, and a Euclidean distance transform $d(x)$ is computed; $d(x)$ provides per–pixel inscribed radii used in all subsequent geometric measurements. Figure~\ref{fig:hrf-open} summarizes the pipeline: panel~(a) shows the raw color fundus image; (b) the binary vessel map from the normalized green channel; and (c) the junction skeleton with per‑junction tariffs and $R(m)$ classes overlaid. In panel~(c), disk RGB encodes the normalized tariff triple $(c_0,c_1,c_2)$ at each degree‑3 junction; the outline color marks the preregistered \emph{R‑band} classes (green: $R<0.55$; amber: $0.55\!\le\!R<0.85$; magenta: $R\!\ge\!0.85$); and a thin white ring denotes junctions where the inferred upkeep exponent $m$ sits at a bracket edge. The tariff disks and bracket‑edge flags are the exact per‑node quantities produced by the pipeline and used downstream; only the \emph{R‑band} colors are interpretive (non‑gating) and do not affect inclusion in the held‑out tests.

The skeleton is converted into a graph under 8–neighborhood connectivity. Degree–3 pixel clusters are merged within 3\,px to define one node per anatomical bifurcation. For each incident branch, unit tangents $\mathbf e_i\in\mathbb R^2$ are estimated by PCA on a fixed arc‑length window (16/12/20\,px for base/lo/hi), with a minimum anisotropy gate $\sigma_1/\sigma_2 \ge 3.0/2.5/3.5$ (base/lo/hi); a chord‑based fallback is used if PCA is unstable. Effective radii $r_i$ are obtained from $d(x)$ by a robust aggregation of distance–transform samples along the branch. The daughter–daughter opening angle is defined as $\theta_{12}=\arccos(\mathbf e_1\!\cdot\!\mathbf e_2)$ and a node is admitted if $\theta_{12}\ge\theta_{\min}$ with $\theta_{\min}\in\{7^\circ,10^\circ,13^\circ\}$ for the lo/base/hi QC suites, and $\theta_{12}\le170^\circ$; \emph{no per‑image auto‑adjustment is applied}. The parent is assigned as the branch most nearly opposite the sum of the other two; near ties are resolved in favor of the branch with higher tangent SVD ratio (tie‑break margin $\tau=0.08$). A conservative cross‑check also evaluates the “fat–daughter as parent’’ assignment, and the worse residual is retained to avoid optimistic bias. This construction yields, at each node, triplets of radii $(r_0,r_1,r_2)$ and directions $(\mathbf e_0,\mathbf e_1,\mathbf e_2)$ together with $\theta_{12}$ and basic per–branch diagnostics, which are encoded visually in panel~(c) of Fig.~\ref{fig:hrf-open}.

Assuming homogeneous tariffs ($c_i\equiv c$, which cancels), the EPIC node law imposes a vector balance
\[
r_0^{m}\,\mathbf e_0 \;+\; r_1^{m}\,\mathbf e_1 \;+\; r_2^{m}\,\mathbf e_2 \;=\; \mathbf 0.
\]
Moving the parent term to the right and squaring norms gives
\[
r_0^{2m} \;=\; \bigl\lVert r_1^{m}\mathbf e_1+r_2^{m}\mathbf e_2 \bigr\rVert^2
= r_1^{2m}+r_2^{2m}+2\cos\theta_{12}\,r_1^{m}r_2^{m},
\]
the scalar “angle–only’’ constraint used to infer $m$ without directions:
\begin{equation}
\label{eq:angle_only_scalar}
r_0^{2m} \;=\; r_1^{2m} + r_2^{2m} + 2\cos\theta_{12}\,r_1^{m}r_2^{m}.
\end{equation}
The vector misclosure is scored by the dimensionless residual
\begin{equation}
\label{eq:closure_residual}
R(m)
=
\frac{\bigl\lVert r_0^{m}\mathbf e_0 + r_1^{m}\mathbf e_1 + r_2^{m}\mathbf e_2 \bigr\rVert_2}{\,r_1^{m}+r_2^{m}\,}\!,
\end{equation}
which is invariant under a global rescaling of radii. The preregistered interpretation zones are $R<0.55$ (``strict'') and $0.55\le R<0.85$ (``loose'').

Two exponents are computed at each node. A \emph{directional} estimate $m_{\text{node}}$ is chosen among three candidates—(i) root solve of Eq.~\eqref{eq:angle_only_scalar}, (ii) a symmetric analytic expression when $r_1\!\approx r_2$, and (iii) a bracketed grid argmin—by selecting the candidate that minimizes $R(m)$ on the measured directions. Exponent solves use the bracket $m\in[0.20,4.00]$; if a root lands at a bracket edge, a single controlled expansion is attempted before falling back to a bracketed grid. A \emph{held–out} estimate $m_{\mathrm{AO}}$ is inferred from the scalar quadruple $(r_0,r_1,r_2,\theta_{12})$ alone using the same solver order; $R(m_{\mathrm{AO}})$ is then evaluated on the measured directions and used for out‑of‑sample tests. Relative tariffs $c=(c_0,c_1,c_2)$ are recovered (up to scale) from the SVD nullspace of $E\,\mathrm{diag}(r^m)$ with gentle nonnegativity clamping, and the associated nullspace residual is recorded as a separate stationarity diagnostic. For the HRF held‑out analyses, \emph{node inclusion is governed by the geometry/QC gates above}. The bands $R(m)<0.55$ (“strict”) and $0.55\le R(m)<0.85$ (“loose”) are \emph{preregistered interpretive zones} and are \emph{not used to gate nodes}. Tariff nullspace fits are diagnostic and \emph{not used for gating}. In HRF (Sato+Otsu), $\sim$23–24\% of nodes land at a bracket edge.

All HRF figures and statistics in this section use the \emph{Sato+Otsu} segmentation together with the pre-specified held-out metric (evaluating $R(m)$ at the angle-only estimate $m_{\mathrm{AO}}$ from $(r_0,r_1,r_2,\theta_{12})$). The HRF reference vessel masks are not used as inputs; all masks are generated algorithmically. The thresholds $0.55$ and $0.85$ for $R(m)$ define interpretive bands used for shading in figures and tables and do not enter any node-selection or quality-control cuts.

\subsection{Results}
\begin{figure}[h]
  \centering
  \includegraphics[width=0.64\columnwidth]{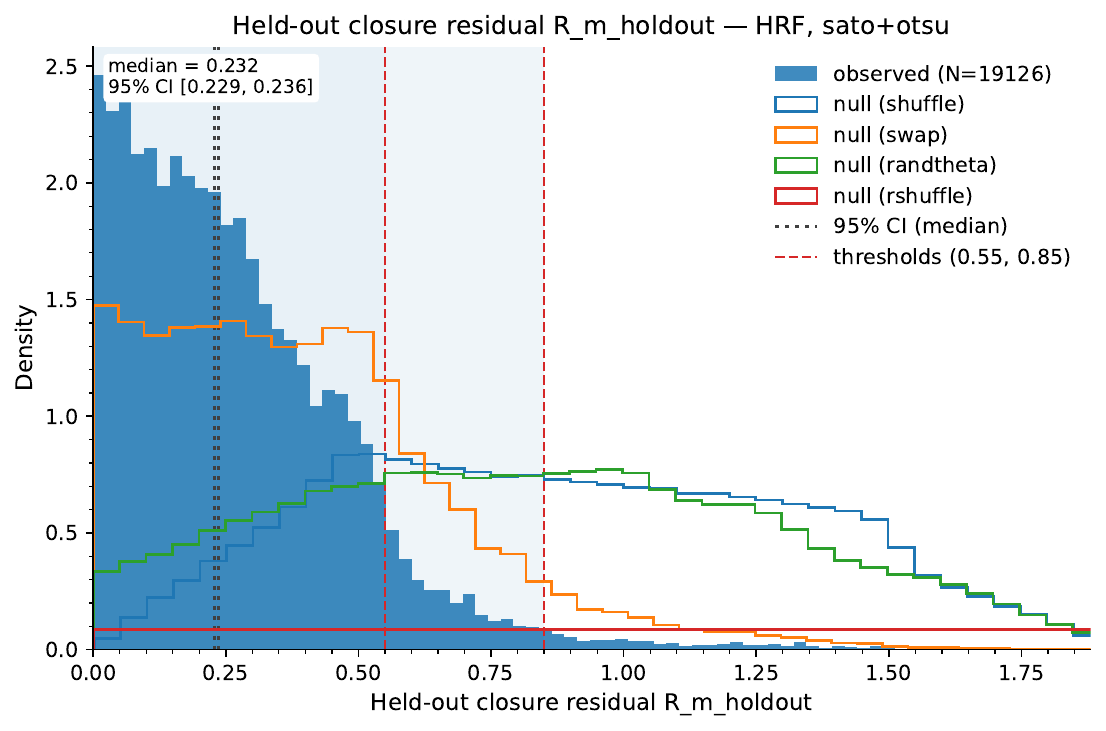}
  \caption{\textbf{Held‑out closure residuals on HRF (\texttt{sato+otsu}).}
  Histogram of $R(m)$ over HRF junctions; the vertical line marks the observed median
  with a 95\% bootstrap CI ($B{=}5000$) \cite{EfronTibshirani1993}. Colored outlines show
  structure‑preserving null controls. Lower values indicate tighter vector closure; shaded bands
  mark preregistered \emph{R‑band (interpretive)} thresholds.}
  \label{fig:residual-hist}
\end{figure}

Figure~\ref{fig:residual-hist} shows the distribution of held–out closure residuals for all degree-3 junctions in HRF segmented with the Sato–Otsu pipeline. For each junction the upkeep exponent \(m\) is first inferred without using the branch directions: the scalar EPIC relation \eqref{eq:angle_only_scalar} is solved from the scalar quadruple \((r_0,r_1,r_2,\theta_{12})\), where \(\theta_{12}\) is the daughter–daughter opening angle, yielding an angle-only estimate \(m_{\mathrm{AO}}\). This exponent is then applied to the measured unit tangents \((\mathbf e_0,\mathbf e_1,\mathbf e_2)\) to compute the held-out closure residual \(R(m_{\mathrm{AO}})\) via Eq.~\eqref{eq:closure_residual}. The filled histogram displays the empirical density of all finite \(R(m_{\mathrm{AO}})\) values, using an adaptive bin count \(\min(80,\max(30,\lceil\sqrt{N}\rceil))\). The sample median is indicated by a vertical line, and a nonparametric 95\% bootstrap confidence interval is shown as dotted vertical lines. Shaded bands mark the preregistered \emph{R‑band (interpretive)} thresholds \(R<0.55\) (strict) and \(0.55\le R<0.85\) (loose); lower residuals correspond to tighter EPIC vector closure at the node.

To test whether small residuals could arise from trivial rearrangements of the data, four structure‑preserving null controls are overlaid as outline density curves, each built from \(n_{\text{perm}}=2000\) permutations (with up to 50 realizations aggregated for display). In the \emph{shuffle} null, the three directions \((\mathbf e_0,\mathbf e_1,\mathbf e_2)\) at each junction are redrawn from that image’s empirical direction pool (falling back to the global pool only if an image has fewer than three directions), while radii \((r_0,r_1,r_2)\) and the exponent \(m\) are held fixed. The \emph{swap} null randomly exchanges the two daughter radii \((r_1,r_2)\) on a node‑by‑node basis, isolating sensitivity to daughter labeling. The \emph{randtheta} null preserves the daughter separation angle \(\theta_{12}\) but randomizes absolute orientation by rotating the daughter pair by a uniform angle \(\phi\); thus \(\mathbf e_1,\mathbf e_2\) become \(\mathbf e_1(\phi),\mathbf e_2(\phi\pm\theta_{12})\) while \(\mathbf e_0\) is unchanged. The \emph{rshuffle} null permutes the triplets \((r_0,r_1,r_2)\) among junctions within each image, leaving directions fixed and thus testing the role of marginal radius statistics. In all four cases, the same angle‑only exponent \(m\) inferred from \((r_0,r_1,r_2,\theta_{12})\) is used when evaluating \(R(m)\), exactly mirroring the held‑out protocol for the observed data.

The resulting pattern is pronounced. The observed distribution (\(N=19{,}126\)) is tightly concentrated below the loose threshold and predominantly within the strict band, with a low median \(\tilde R=0.232\) and 95\% bootstrap interval \([0.228,0.236]\), and approximately \(91\%\) of nodes satisfying \(R(m)<0.55\). All four nulls are displaced to larger \(R(m)\), exhibiting substantially poorer closure once local geometric couplings are disrupted by direction resampling, daughter relabeling, absolute‑orientation randomization, or radius shuffling. Because \(m\) is estimated from \((r_0,r_1,r_2,\theta_{12})\) alone and then applied to held‑out directions, the clear separation between the filled histogram and every outlined null indicates genuine, out‑of‑sample vector closure consistent with the EPIC node law, rather than artifacts of orientation pools, labeling conventions, absolute orientation, or marginal radius structure.

\begin{figure}[t!]
  \centering
  \includegraphics[width=0.64\columnwidth]{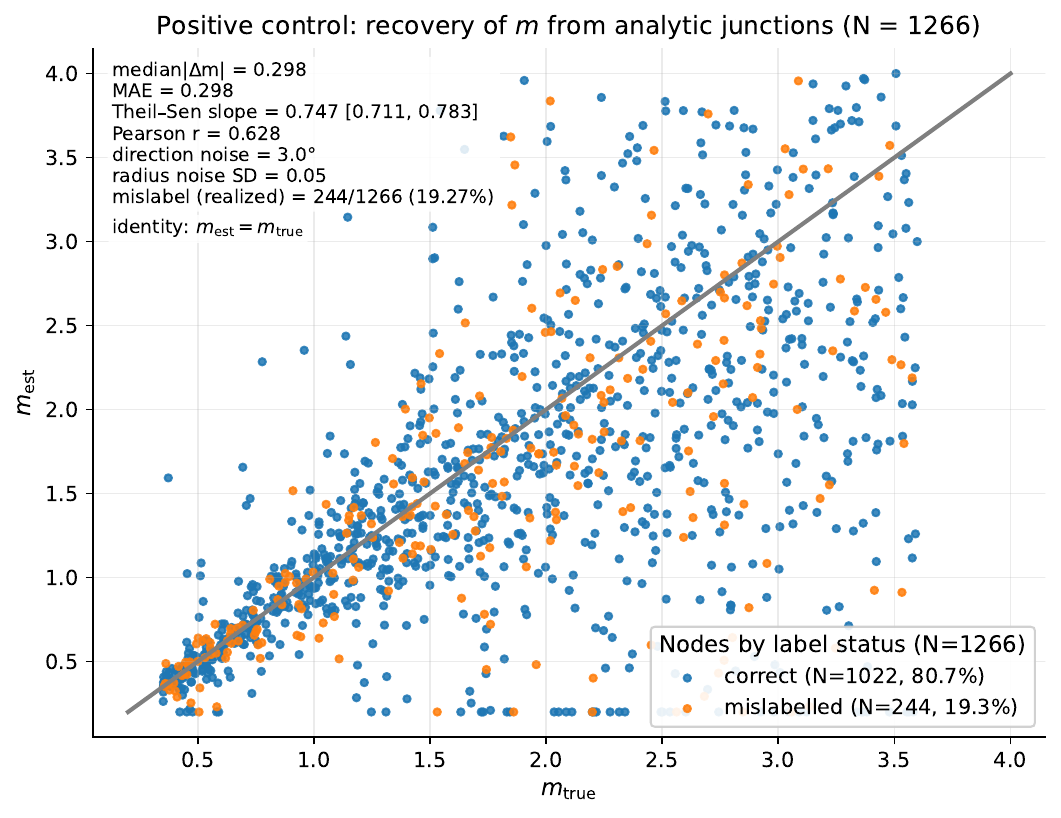}
  \caption{\textbf{Positive control: recovery of $m$ from analytic junctions.}
  Estimated upkeep exponents $m_{\mathrm{est}}$ versus ground‑truth $m_{\mathrm{true}}$
  for $N=1266$ synthetic Y‑junctions with directional and radius noise and a
  $20\%$ parent‑mislabel fraction. Blue points denote correctly labelled nodes; orange
  points denote mislabelled nodes. The grey line shows the identity $m_{\mathrm{est}}=m_{\mathrm{true}}$.}
  \label{fig:posctrl-recovery}
\end{figure}

Figure~\ref{fig:posctrl-recovery} provides a positive control for the angle‑only inversion used for HRF bifurcations. Synthetic Y‑junctions are generated with known exponents \(m_{\mathrm{true}}\), radii \(r_1,r_2\in[1,4]\), and opening angles \(\theta_{12}\in[15^\circ,160^\circ]\); parent radii and directions are then constructed so that the scalar EPIC relation holds exactly. A deliberately conservative noise model is imposed: radii are perturbed by independent Gaussian factors with 5\% standard deviation, branch directions are rotated by 3\(^\circ\) RMS, and in 20\% of cases the parent is forcibly swapped with the thicker daughter. From each noisy quadruple \((r_0,r_1,r_2,\theta_{12})\), the exponent \(m\) is re‑estimated on the physical bracket and compared to \(m_{\mathrm{true}}\).

The resulting scatter of \(m_{\mathrm{est}}\) versus \(m_{\mathrm{true}}\) shows that the inversion retains substantial information about the underlying scaling law. Correctly labelled junctions (blue) form a tight, monotone cloud around the identity, with median absolute error \(|\Delta m|\simeq 0.30\), Theil–Sen slope \(0.747\,[0.711,0.783]\), and Pearson correlation \(r=0.628\). Mislabelled junctions (orange) fall systematically below the diagonal, reflecting the expected conservative bias when a daughter branch is treated as the parent and the inferred exponent is driven downward. Under a pessimistic but anatomically motivated mix of noise and mislabelling, the angle‑only estimator remains quantitatively informative while exhibiting a well‑characterized downward bias, supporting the interpretation that strong held‑out closure at real retinal junctions is unlikely to be an artifact of the inversion procedure and instead reflects genuine geometric consistency with the EPIC node law.

Overall, the first‑pass analysis is summarized in Table~\ref{tab:HRF_EPIC_summary}, which compiles held‑out closure statistics and robustness diagnostics for HRF. Central values are taken from the Sato–Otsu segmentation core in the preregistered held‑out configuration, evaluated under a matched base/lo/hi QC triplet. In the \emph{base} setting the daughter–daughter angle gate is \(10^\circ\), the tangent is fit over 16\,px, and the minimum PCA SVD ratio is \(3.0\) (i.e., \(\sigma_1/\sigma_2\)); the \emph{lo} variant relaxes these thresholds to \(7^\circ\), 12\,px, and \(2.5\); and the \emph{hi} variant tightens them to \(13^\circ\), 20\,px, and \(3.5\). For each median quoted in the table, the suite‑level systematic uncertainty is obtained by taking the half‑range of that median across these three QC settings and combining it in quadrature with the corresponding half‑range across the HRF segmentation triplet; the result is reported as the “syst’’ term accompanying the bootstrap interval.

The rows are interpreted as follows. \(N_{\rm nodes}\) is the total number of degree‑3 junctions that survive all geometric and QC cuts in the base configuration. The “Node–median’’ and “Image–median’’ lines report, respectively, the median of \(R(m)\) over all nodes and the median of per‑image medians; each is accompanied by a nonparametric 95\% bootstrap interval (subscript “stat’’) and the suite‑level systematic just described (subscript “syst’’). “Nodes with \(R(m)<0.55\)’’ gives the fraction of junctions in the preregistered strict band. The “Shuffle–null median’’ is the median \(R(m)\) under the strongest structure‑preserving null (direction shuffle), and Cliff’s \(\delta\) quantifies the effect size between the observed and shuffle‑null \(R(m)\) distributions (negative values indicate tighter closure in the data, with \(|\delta|\) in the “large’’ regime). The “Worst null \(p\)’’ entries take, for both the median and the strict‑band fraction, the largest one‑sided permutation \(p\) across the four preregistered null controls (resolution‑limited at \(\approx 5\times 10^{-4}\) for \(n_{\text{perm}}{=}2000\)), demonstrating that all nulls are rejected at the stated level. Finally, the ablation rows summarize the range of node‑median \(R(m)\) obtained when the tangent SVD gate and minimum angle are scanned over the robustness grid, together with the worst permutation \(p\) across those cells and the maximum change in the strict‑QC pass fraction. The narrow spread in medians (0.221–0.226), uniformly small \(p\)-values, and sub‑percent drift in the strict‑QC fraction jointly indicate that the closure signal is stable under these geometry and QC perturbations and, relative to the base median (0.232), is slightly tightened under ablations.

\begin{table}[t]
  \centering
  \footnotesize
  \setlength{\tabcolsep}{6pt}
  \renewcommand{\arraystretch}{1.15}
  \begin{tabular}{ll}
    \toprule
    \multicolumn{2}{c}{\textbf{HRF (sato+otsu core)}}\\
    \midrule
    $N_{\rm nodes}$                   & $19\,126$ \\
    \addlinespace[2pt]
    Node-median $R(m)$                & $0.232\;[0.228,\,0.236]_{\rm stat}\ \pm 0.020_{\rm syst}$ \\
    Image-median $R(m)$               & $0.227\;[0.220,\,0.237]_{\rm stat}\ \pm 0.021_{\rm syst}$ \\
    Nodes with $R<0.55$ (strict band) & $91.2\%$ \\
    \addlinespace[2pt]
    Shuffle-null median $R(m)$        & $0.872$ \\
    Cliff’s $\delta$ (data vs shuffle) & $-0.83$ (large) \\
    Worst null $p$ (median $R$)       & $\le 5\times10^{-4}$ \\
    Worst null $p$ (frac.\ $R<0.55$)  & $\le 5\times10^{-4}$ \\
    \addlinespace[2pt]
    Ablation median $R(m)$ range      & $0.221\text{--}0.226$ \\
    Ablation worst $p$                & $5\times10^{-4}$ \\
    Strict-QC drift across ablations  & $0.9$ percentage points \\
    \bottomrule
  \end{tabular}
  \caption{\textbf{Held‑out closure on HRF and associated uncertainties.}
  Node‑ and image‑median residuals $R(m)$ are reported for the base HRF configuration, with nonparametric 95\% bootstrap intervals (“stat”) and suite‑level systematics (“syst”); see text for the definition of the QC triplet and the systematic construction. The remaining rows summarize the strongest structure‑preserving null (shuffle), the corresponding effect size and $p$‑values (resolution‑limited by $n_{\text{perm}}{=}2000$), and the stability of $R(m)$ under the SVD/angle ablation grid.}
  \label{tab:HRF_EPIC_summary}
\end{table}

\subsection{Conclusions and future steps}
The HRF experiment provides a first, preregistered demonstration of \emph{first–type closure}: when the upkeep exponent \(m\) is inferred from scalar geometry alone \((r_0,r_1,r_2,\theta_{12})\), the held–out vector residuals are sharply left–shifted and robust to segmentation/QC scans, while structure–preserving nulls move decisively right. In this way the HRF analysis implements a single falsifiable pathway from EPIC to data: the upkeep exponent \(m\) is inferred from scalar geometry \((r_0,r_1,r_2,\theta_{12})\), and the resulting prediction is then tested by vector closure on the measured branch directions. The next step is to deliberately seek both confirmations and failures of this closure in settings where EPIC makes distinct, testable predictions (with the underlying formalism developed in \S\ref{sec:epic_optics} and in the Growth–cone section, proof in Appendix~\ref{sec:cone_proof}):
\begin{enumerate}
  \item \textbf{Independent retinal corpora and modalities.} Repeat the held–out closure test on DRIVE, STARE, CHASE–DB1, OCT–A, and fluorescein angiography; report \(R(m)\) medians, strict–band fractions, and effect sizes against preregistered nulls across datasets and resolutions.
  \item \textbf{Three–dimensional vasculature.} Apply the same protocol to 3D centerlines from cleared–tissue light–sheet or \(\mu\)CT volumes (brain, kidney, lung); test 3D vector closure and the symmetric–angle predictions, including the mixed surface/volume weighting in Eq.~\eqref{eq:angle_mixed}.
  \item \textbf{Other organs and scales.} Probe cerebral, coronary, renal, and pulmonary beds; compare inferred \(m\), check the priced–shear invariant \(\mathscr S_i\), and relate deviations to physiological differences (loops, collateralization).
  \item \textbf{Rheology–aware checks.} Where apparent viscosity varies with radius \(\mu(r)\), test the predicted shift in the flow–radius slope \(\alpha_{\rm eff}(r)\) and the constancy of \(\mathscr S_i\) across branches (Appendix~\ref{app:rheology}).
  \item \textbf{Refraction (“EPIC optics”) in phantoms.} In microfluidic Y–arrays, impose controlled steps/gradients in the composite price field \(\big[\kappa_{\mathrm{hyd}}^{\,m}(\varepsilon_b c)^{\,n}\big]^{1/(m+n)}\) and measure Snell–type refraction; estimate \(\mathfrak n_2/\mathfrak n_1\) from angles and compare to the engineered contrast (formalism in \S\ref{sec:epic_optics}, cf.\ Eqs.~\eqref{eq:tariff_index}–\eqref{eq:Snell}).
  \item \textbf{Equipartition audits.} Where control–layer power is measurable, test the dilation identity \(nP_{\rm flow}=mP_{\rm struct}\) and the predicted power split \(m:(m{+}n)\) at steady operation (\S\ref{sec:epic_optics}).
  \item \textbf{Concavity and loop statistics.} Quantify tree depth and loop prevalence; test subadditivity implied by \(\gamma=2m/(m{+}n)<1\) by comparing costs of merged vs.\ isolated routes in data and phantoms, and by examining fluctuation/damage correlates of looping (\S\ref{sec:epic_optics}).
  \item \textbf{Tariff tomography.} Infer relative node tariffs \((c_1/c_0,c_2/c_0)\) from geometry (Appendix~\ref{app:tomography}) and ask whether they shift across health vs.\ disease strata or along arterial–venous trees; cross–check against independent proxies where available.
  \item \textbf{Growth–cone bound (time–resolved falsification).} In angiogenesis or engineered networks with metered power and boundary signaling, test \(\dot I\le \min(P_{\rm flow}/\varepsilon_b,\ \int_{\Sigma}\mathcal C_\epsilon\,dA)\) by jointly estimating both faces of the bound (definition in Eq.~\eqref{eq:EPIC_cone_node}; proof in Appendix~\ref{sec:cone_proof}). A crisp falsification attempt is to \emph{toggle the limiting face}: hold \(P_{\rm flow}\) roughly fixed while modulating boundary capacity, and vice versa—the measured \(\dot I\) should track the smaller budget with the predicted crossover.
  \item \textbf{Outside vascular biology.} Examine branched media with clear junctions—leaf venation, dendritic crystal (snowflake) triple lines, slime–mold and river confluences—for angle closure using the appropriate geometric tariff (surface vs.\ area) as a generality stress test.
\end{enumerate}
Together, these studies turn EPIC’s geometric statements—closure, weighted Murray scaling, refraction, equipartition, concavity, and growth limits—into a suite of falsifiable targets across modalities, organs, and platforms, with the formal machinery developed in \S\ref{sec:epic_optics} and the Growth–cone section (proof in Appendix~\ref{sec:cone_proof}).

\section{Growth‑cone bound: rate limits for reliable structure}
The EPIC ledger imposes a real‑time, unit‑consistent constraint on the creation of \emph{reliably decodable, stably maintained distinctions} within any space–time control volume. At each instant the achievable rate is limited by the smaller of two budgets: an interior budget purchased from dissipation at the audited tariff and a boundary budget set by the reliably decodable information that can traverse the enclosing surface. This two‑face constraint—the \emph{growth–cone} bound—rests on nonequilibrium entropy production, an operational Landauer price, and capacity‑limited boundary signaling; the formal proof appears in Appendix~\ref{sec:cone_proof}.

Consider a control volume \(V\subset\mathbb{R}^{3}\) with (possibly time‑dependent) boundary \(\Sigma_t\) and outward normal \(\mathbf n\). The audit fixes the effective energetic price per reliable bit as \(\varepsilon_b(x,t)=\zeta(x,t)\,k_B T(x,t)\ln 2\) with \(\zeta(x,t)\ge 1\), and thereby defines the power‑priced information–throughput density \(\Psi_b(x,t)\equiv \sigma_s(x,t)\,T(x,t)/\varepsilon_b(x,t)\) \([\si{\bit\per\metre\cubed\per\second}]\). On the boundary, let \(\mathcal C_\epsilon(N;x,t)\) denote the \(\epsilon\)–reliable local rate density \([\si{bit\,m^{-2}\,s^{-1}}]\) supported by the available physical channels operated at blocklength \(N\) (stochastic‑thermodynamics sense). Writing the stock of reliably maintained distinctions as \(I(t)=\int_V \iota(x,t)\,dV\) \([\si{bit}]\), the instantaneous creation rate obeys
\begin{equation}
\label{eq:EPIC_cone}
\frac{d}{dt}\!\int_{V}\iota\,dV
\;\le\;
\min\!\left\{
\int_{V}\Psi_b(x,t)\,dV\ ,\
\int_{\Sigma_t}\! \mathcal C_\epsilon(N;x,t)\,dA
\right\}.
\end{equation}
The interior face purchases reliable bits from entropy production at the audited tariff, while the boundary face limits the influx of genuinely new information through \(\Sigma_t\). Equality can hold only in the ideal limit where the nonlimiting face (interior or boundary) is truly slack and the limiting face operates reversibly at the audited reliability, i.e., every watt or channel use is fully devoted to reliably coded bits with no wasted margin. Any additional finite-time overheads, redundancy or proofreading, idle duty cycles, or unmetered leakage channels consume part of the energetic or communication budget and therefore lower the actually attainable rate below the growth–cone bound.

In a steady, isothermal node the interior integral simplifies to \(\int_V \Psi_b\,dV=\int_V \sigma_s T\,dV/\varepsilon_b=P_{\rm flow}/\varepsilon_b\), so the bound reduces to
\begin{equation}
\label{eq:EPIC_cone_node}
\dot I
\;\le\;
\min\!\left\{
\frac{P_{\rm flow}}{\varepsilon_b}\ ,\
\int_{\Sigma_t}\! \mathcal C_\epsilon(N;x,t)\,dA
\right\}.
\end{equation}
Interpreted in informational units, the first term states that, at a fixed tariff, each interior watt can fund at most \(\varepsilon_b^{-1}\) reliable \si{bit/s}, while the second term caps the influx of new reliable bits by the boundary’s communication capacity at the chosen blocklength \(N\) and error probability \(\epsilon\). When boundary signaling admits an information–theoretic description with per–use capacity \(C\) (bit/use) and dispersion \(V\), operated at symbol rate \(W(x,t)\) (use/s) across a channel density \(\rho_{\rm ch}(x,t)\) (m\(^{-2}\)), the normal approximation gives
\[
\mathcal C_\epsilon(N;x,t)
=\rho_{\rm ch}(x,t)\,W(x,t)\,
\Big[\,C-\sqrt{V/N}\,\Phi^{-1}(\epsilon)\,\Big]_+,
\]
where the bracketed term is the finite–blocklength penalty (vanishing as \(N^{-1/2}\)) and the prefactor \(\rho_{\rm ch}W\) converts per–use rates into a per–area, per–time density, so that \(\int_{\Sigma_t}\mathcal C_\epsilon\,dA\) has units \si{bit/s}. On the interior side, \(\Psi_b=\sigma_s T/\varepsilon_b\) prices \si{W\,m^{-3}} into \si{bit\,m^{-3}\,s^{-1}}, so \(\int_V \Psi_b\,dV\) likewise carries units \si{bit/s}. The two faces of \eqref{eq:EPIC_cone} therefore live in a common currency, and the growth–cone can be viewed as the domain–level analogue of the node objective: interior dissipation and boundary signaling are entered on the same ledger and can be compared directly—under heterogeneous, space- and time-dependent operating conditions—in units of reliably decodable \si{bit/s}. Relation to thermodynamic uncertainty relations is discussed in Appendix~\ref{app:TUR_remark}.

\section{Design corollaries: equipartition, concavity, refraction}
\label{sec:epic_optics}
This section adopts a single, explicit set of assumptions: steady flow with fixed demands, and per–length contributions homogeneous in radius—transport \(Q^2 r^{-n}\) and upkeep \(r^{m}\) with \(m,n>0\)—so that Euler’s theorem for homogeneous functions and stationarity of \(\mathcal L\) (Eq.~\eqref{eq:EPIC_additive_node}) can be applied without further constitutive detail. In this form, the derivations lie squarely in the classical minimum–work and optimal–network lineage of Murray, Sherman, and Zamir \cite{Murray1926a,Murray1926b,Sherman1981,Zamir1976Optimality}, and in their efficient–network extensions in physics and biology where scaling laws and network architecture emerge from variational arguments \cite{Banavar1999Nature,WestBrownEnquist1997Science,WestBrownEnquist1999Nature}. The same homogeneity structure underpins concave–cost and branched–transport theory, in which subadditive flux costs generically favor mergers and tree–like hierarchies while robustness and variability promote loops \cite{Gilbert1967,BernotCasellesMorel2009,BohnMagnasco2007PRL,Katifori2010PRL}. Viewing routing in heterogeneous media as a Fermat problem once a dimensionless index is specified then links the present construction directly to the optics and Hamilton–Jacobi perspective on geodesics in weighted media \cite{BornWolf1999,SethianVladimirsky2003,BaoChernShen2000}.

EPIC contributes three design–level statements that are used repeatedly in what follows. First, a uniform dilation of all radii, applied to per–length contributions homogeneous of degrees \(-n\) and \(m\), yields an audit–testable equipartition identity \(nP_{\rm flow}=mP_{\rm struct}\) [Eq.~\eqref{eq:equipartition}], fixing the optimal power split \(P_{\rm flow}:(P_{\rm flow}+P_{\rm struct})=m:(m+n)\) independently of topology or load. Second, eliminating the radii at stationarity produces a concave, flux–only per–length edge cost \(\mathcal{C}^\star(Q;x)\propto \kappa_{\mathrm{hyd}}(x)^{\frac{m}{m+n}}[\varepsilon_b(x)c(x)]^{\frac{n}{m+n}} Q^{\gamma}\) with exponent \(\gamma=2m/(m+n)\) [Eq.~\eqref{eq:tariff_index}], so that in the regimes of interest \(m<n\) one has \(0<\gamma<1\) and merging flows into shared trunks is energetically preferred over isolating them. Third, factoring the spatial dependence in \(\mathcal{C}^\star\) defines a \emph{dimensionless} routing index \(\mathfrak n(x)\) [Eq.~\eqref{eq:EPIC_index_def}] that recasts least–cost routing as an “optical path length’’ problem \(\int \mathfrak n(x)\,ds\) with a Snell–type refraction law across tariff contrasts [Eqs.~\eqref{eq:EPIC_optical_length}, \eqref{eq:Snell}]. Subsec.~\ref{sec:epic_optics}A derives the dilation identity, Subsec.~\ref{sec:epic_optics}B develops the concave edge cost and its exponent \(\gamma\), and Subsec.~\ref{sec:epic_optics}C introduces \(\mathfrak n(x)\) and the associated refraction conditions.

\subsection*{A. Equipartition under dilation}

EPIC adopts the same power–law forms used in optimal–network theory: per–segment transport contribution \(P_{{\rm flow},i}=\kappa_i Q_i^2 r_i^{-n}\) with \(n>0\) (Poiseuille \(n=4\) for Newtonian laminar tubes; other \(n\) capture slip or non‑Newtonian effects), and a structural pricing law \(\mathcal B_i=c_i r_i^{m}\) with \(m>0\) (volume–priced \(m=2\), surface–priced \(m=1\), or a calibrated mix) \cite{WhiteViscous,BruusMicrofluidics,BirdStewartLightfoot2007}. The coefficients \(\kappa_i,c_i>0\) may absorb length and material factors but are independent of \(r_i\). For fixed topology and exogenous demands \(Q_i\), the node‑level ledger
\[
\mathcal L(\{r_i\})=\sum_i \kappa_i Q_i^2 r_i^{-n} \;+\; \Lambda \sum_i c_i r_i^{m}
\;\equiv\; P_{\rm flow}+P_{\rm struct}
\]
is examined at interior optima (all \(r_i>0\)).

Because the two terms are homogeneous in the radii (degrees \(-n\) and \(m\), respectively), a uniform dilation \(r_i\!\mapsto\!(1{+}\epsilon)r_i\) and Euler’s theorem immediately give the \emph{equipartition} identity
\begin{equation}
\label{eq:equipartition}
n\,P_{\rm flow}=m\,P_{\rm struct},
\end{equation}
so the optimal power split is fixed by the exponents alone:
\[
\frac{P_{\rm flow}}{P_{\rm flow}+P_{\rm struct}}=\frac{m}{m+n},
\qquad
\frac{P_{\rm struct}}{P_{\rm flow}+P_{\rm struct}}=\frac{n}{m+n}.
\]
Edgewise stationarity enforces the local balance \(n\,\kappa_i Q_i^2 r_i^{-n}=m\,\Lambda c_i r_i^{m}\), which yields the optimal radius scaling
\[
r_i^{\,m+n}=\frac{n\,\kappa_i}{m\,\Lambda\,c_i}\,Q_i^2
\quad\Longrightarrow\quad
r_i\propto \Lambda^{-1/(m+n)}\,Q_i^{\,2/(m+n)}.
\]
Thus a global tariff rescaling \(\Lambda\mapsto a\Lambda\) shrinks all optimal radii by the common factor \(a^{-1/(m+n)}\) without changing either the power split (\eqref{eq:equipartition}) or the branch angles, which are fixed by translation stationarity (Eq.~\eqref{eq:angle_vector}) and do not depend on \(\Lambda\).

Physically, \((m,n)\) alone set how an optimal network allocates its budget between transport and structure. For the canonical Newtonian/volume case \((n,m)=(4,2)\), \(P_{\rm flow}:P_{\rm struct}=1:2\); for Newtonian/surface \((4,1)\), \(1:4\) \cite{Murray1926a,Sherman1981,Zamir1976Optimality}. These ratios are audit targets: systematic deviations indicate active constraints, departures from the assumed homogeneities (e.g., radius‑dependent effective \(n\) from Fåhræus–Lindqvist or shear‑thinning) \cite{Pries1992,Secomb2013,Ascolese2019}, or operation away from stationarity. Practically, one can compare measured \(P_{\rm flow}=\sum_i Q_i\Delta p_i\) and \(P_{\rm struct}=\Lambda\sum_i c_i r_i^m\) (with \(\varepsilon_b\) calibrated) to the predicted ratio \(m:n\). The same homogeneity structure that yields \eqref{eq:equipartition} also leads to a strictly concave effective edge cost with exponent \(\gamma=2m/(m{+}n)\) (Sec.~\ref{sec:epic_optics}B), recovering the energetic preference for mergers and deep trees emphasized in PRL‑level analyses \cite{BohnMagnasco2007PRL,Katifori2010PRL}.

\subsection*{B. Concavity of the effective edge cost and the prevalence of trees}

Classical minimum–work analyses of vascular design balance a per–length pumping term against a per–length maintenance term, eliminate the radius at stationarity, and obtain an effective edge cost that depends only on flux \cite{Murray1926a,Murray1926b,Sherman1981,Zamir1976Optimality}. PRE/PRL studies have shown that when both contributions are homogeneous in \(r\), the eliminated cost scales as a power of \(Q\), and that a sublinear exponent makes this cost strictly concave so that tree–like hierarchies are energetically favored \cite{Durand2006PRE,Durand2007PRL,BohnMagnasco2007PRL,Katifori2010PRL}. EPIC retains this homogeneity structure but prices upkeep in calibrated units (J/bit), so the same elimination yields not only the concavity exponent but also its explicit dependence on the informational tariff.

At a fixed location \(x\), consider a single conduit of radius \(r\) carrying flow \(Q\) in a medium with local tariff \(\varepsilon_b(x)c(x)\). Under the same homogeneity assumptions used above—per–length transport dissipation \(P_{{\rm flow},i}\propto Q_i^2 r_i^{-n}\) and structural pricing \(\mathcal B_i\propto r_i^{m}\) with \(m,n>0\)—its per–length contribution to the ledger (up to geometry–independent constants) can be written as
\[
\mathcal{C}(Q,r;x)=\kappa_{\mathrm{hyd}}(x)\,Q^{2}r^{-n}+\Lambda\,c(x)\,r^{m},
\]
with \(\kappa_{\mathrm{hyd}}(x),c(x)>0\). Here \(\kappa_{\mathrm{hyd}}(x)\) encodes hydraulic factors (viscosity, length, roughness), \(c(x)\) encodes local structural bit–density and duty cycle, and \(\Lambda=\lambda\,\varepsilon_b\) prices each structural bit in \si{J/bit}. Minimizing \(\mathcal{C}(Q,r;x)\) with respect to \(r>0\) at fixed \(Q\) eliminates the radius and yields an optimal per–length cost of the form
\begin{equation}
\label{eq:tariff_index}
\mathcal{C}^\star(Q;x)\;\propto\;
\kappa_{\mathrm{hyd}}(x)^{\frac{m}{m+n}}\,
\bigl[\varepsilon_b(x)c(x)\bigr]^{\frac{n}{m+n}}\,
Q^{\gamma},
\qquad
\gamma\equiv\frac{2m}{m+n}.
\end{equation}
The factor \([\varepsilon_b(x)c(x)]^{n/(m+n)}\) is a purely tariff–dependent prefactor (“price field’’), while the exponent \(\gamma\in(0,2)\) quantifies how sharply cost grows with flux. In the regimes of interest, with structural pricing shallower than the hydraulic exponent (e.g.\ Poiseuille \(n=4\) and \(m\in\{1,2\}\)), 
\[
0<\gamma=\frac{2m}{m+n}<1,
\]
so the flux–only edge cost is strictly concave.

For \(0<\gamma<1\), the map \(Q\mapsto Q^{\gamma}\) is strictly concave, and hence
\[
\mathcal{C}^\star(Q_1{+}Q_2;x)\;<\;\mathcal{C}^\star(Q_1;x)+\mathcal{C}^\star(Q_2;x)
\qquad(Q_1,Q_2>0).
\]
That is, combining two demands into a shared trunk is strictly cheaper than carrying them on isolated paths. This strict subadditivity of the optimal edge cost provides a direct, tariff–aware explanation for the empirical prevalence and depth of tree–like hierarchies in optimal transport networks \cite{BohnMagnasco2007PRL,Katifori2010PRL}.

The exponent \(\gamma\) quantifies the strength of this effect. Smaller \(\gamma\) (more concave cost; e.g., surface–priced upkeep with \(m=1\) or reduced effective \(n\) from slip or shear–thinning) favors early mergers and thicker upstream trunks; larger \(\gamma\) (less concave cost) delays mergers and yields shallower trees. These architectural trends are consistent with constructal arguments for tree–shaped flow architectures \cite{BejanLorente2004,Bejan2010} and with vascular optimality analyses \cite{Zamir1976Optimality}. Departures from purely tree–like structures—loops—then arise naturally when fluctuations, damage, or robustness pressures are added on top of this deterministic concave baseline \cite{Corson2010PRL,Katifori2010PRL}.

\subsection*{C. Routing index and Snell-like refraction (“EPIC optics”)}

The effective edge cost \eqref{eq:tariff_index} cleanly separates the dependence on the carried flux \(Q\) from the dependence on the local medium and tariff. For a short segment centered at position \(x\), the per–length contribution to the EPIC ledger can be written as \(\mathcal{C}(Q,r;x)=\kappa_{\mathrm{hyd}}(x)\,Q^{2}r^{-n}+\Lambda\,c(x)\,r^{m}\) with \(m,n>0\). Here \(\kappa_{\mathrm{hyd}}(x)\) is the hydraulic prefactor per unit length (for Poiseuille flow, \(n{=}4\) and \(\kappa_{\mathrm{hyd}}\!\propto\!\mu\)), \(c(x)\) encodes the local structural bit–density per unit geometry, and \(\Lambda=\lambda\,\varepsilon_b(x)\) prices reliable bits in \si{J/bit}.

Minimizing \(\mathcal{C}\) with respect to \(r>0\) at fixed \(Q\) yields the optimal per–length cost at the stationary radius \(r^\star\):
\begin{equation}
\label{eq:edge_cost_local_fixedQ}
\mathcal{C}^{\star}(Q;x)\ \propto\
\kappa_{\mathrm{hyd}}(x)^{\frac{m}{m+n}}\,[\varepsilon_b(x)\,c(x)]^{\frac{n}{m+n}}\;Q^{\gamma},
\qquad
\gamma\equiv \frac{2m}{m+n}.
\end{equation}
Thus, for a given flux \(Q\), all spatial variation enters through the composite factor
\[
\kappa_{\mathrm{hyd}}(x)^{\frac{m}{m+n}}[\varepsilon_b(x)\,c(x)]^{\frac{n}{m+n}}.
\]

It is therefore convenient to factor the fixed–\(Q\) cost into a purely flow–dependent term \(Q^{\gamma}\) and a purely “medium\,+\,tariff’’ term. With \(\Lambda=\lambda\,\varepsilon_b(x)\), the \emph{EPIC index} is defined as
\begin{equation}
\label{eq:EPIC_index_def}
\mathfrak n(x)\ \equiv\
\left[\frac{\kappa_{\mathrm{hyd}}(x)}{\kappa_{0}}\right]^{\frac{m}{m+n}}
\left[\frac{\varepsilon_b(x)\,c(x)}{\varepsilon_{b0}\,c_{0}}\right]^{\frac{n}{m+n}},
\end{equation}
a dimensionless composite price field unique up to an overall multiplicative constant. The reference scales \(\kappa_{0}\) and \(\varepsilon_{b0}c_{0}\) determine only that overall factor and therefore do not affect geodesics or the Snell-like refraction law derived below. Along any candidate route \(\Gamma\) on which the flux \(Q\) is fixed (single–commodity routing), the total priced cost is
\[
\mathcal J[\Gamma]\ \propto\ \int_{\Gamma}\mathcal{C}^\star(Q;x)\,ds
\;=\;Q^{\gamma}\!\int_{\Gamma}\mathfrak n(x)\,ds .
\]
At fixed \(Q\), choosing a least–cost path is therefore equivalent to minimizing the “optical path length’’
\begin{equation}
\label{eq:EPIC_optical_length}
\mathcal S[\Gamma]\ \equiv\ \int_{\Gamma}\mathfrak n(x)\,ds,
\end{equation}
in direct analogy with geometric optics, where rays minimize \(\int n_{\mathrm{opt}}(x)\,ds\) in a medium of refractive index \(n_{\mathrm{opt}}(x)\)~\cite{BornWolf1999,LandauLifshitzEDCM}. When \(\kappa_{\mathrm{hyd}}\) is spatially uniform, \eqref{eq:EPIC_index_def} reduces to \(\mathfrak n(x)=[\varepsilon_b(x)c(x)/(\varepsilon_{b0}c_{0})]^{\,n/(m+n)}\).

Two standard consequences follow. For a sharp interface between regions of constant index \(\mathfrak n_1\) and \(\mathfrak n_2\), varying the crossing point while holding endpoints fixed yields the Snell-type condition
\begin{equation}
\label{eq:Snell}
\mathfrak n_1\,\sin\theta_1\;=\;\mathfrak n_2\,\sin\theta_2,
\end{equation}
where \(\theta_i\) is the angle between the path and the interface normal in region \(i\). Entering a cheaper region (smaller \(\mathfrak n\)) bends the stationary route away from the normal; entering a more expensive region bends it toward the normal. When \(\mathfrak n_1>\mathfrak n_2\) and \(\sin\theta_1>\mathfrak n_2/\mathfrak n_1\), no transmitted stationary path exists and the optimal route undergoes “total internal reflection’’ at the interface.

In smoothly varying media, the Euler–Lagrange equation for \eqref{eq:EPIC_optical_length} yields the ray equation
\[
\frac{d}{ds}\bigl(\mathfrak n\,\hat{\mathbf t}\bigr)\;=\;\nabla_\perp \mathfrak n,
\]
where \(\hat{\mathbf t}\) is the unit tangent to the path and \(\nabla_\perp\) denotes the component of \(\nabla\mathfrak n\) perpendicular to \(\hat{\mathbf t}\). Stationary routes curve toward decreasing \(\mathfrak n\), i.e., toward cheaper composite conditions set jointly by hydraulics \(\kappa_{\mathrm{hyd}}(x)\) and tariffs \(\varepsilon_b(x)c(x)\). In anisotropic or heterogeneous media—where any of \(\kappa_{\mathrm{hyd}}(x)\), \(\varepsilon_b(x)\), or \(c(x)\) varies in space—conduits preferentially “hug’’ low-\(\mathfrak n(x)\) corridors, and at the network scale this produces characteristic funneling patterns that can be read directly from maps of \(\mathfrak n(x)\).

Because \(\mathfrak n(x)=\big[\kappa_{\mathrm{hyd}}(x)/\kappa_{0}\big]^{\frac{m}{m+n}}\big[\varepsilon_b(x)c(x)/(\varepsilon_{b0}c_{0})\big]^{\frac{n}{m+n}}\), independent estimates of the hydraulic prefactor \(\kappa_{\mathrm{hyd}}(x)\), the energetic tariff \(\varepsilon_b(x)\) (audit–level J/bit), and the structural coefficient \(c(x)\) (bit–density per geometry) convert observed routing patterns into a quantitative \(\mathfrak n(x)\) field. A particularly clean falsification test is a tissue–engineered or microfluidic system with a controlled step in the composite field \(\big[\kappa_{\mathrm{hyd}}^{\,m}(\varepsilon_b c)^{n}\big]^{1/(m+n)}\) across a planar interface (or simply in \([\varepsilon_b c]\) if \(\kappa_{\mathrm{hyd}}\) is uniform): measured incidence and refraction angles \((\theta_1,\theta_2)\) must then satisfy \eqref{eq:Snell}. When direct audits are not feasible, geometry still constrains \emph{ratios}: \eqref{eq:Snell} fixes \(\mathfrak n_2/\mathfrak n_1\) across sharp transitions, and the curvature of stationary paths constrains \(\nabla \mathfrak n(x)\). In both regimes, refraction signatures provide sharp, testable links between spatial variation in the composite price field and the observed geometry of transport routes.

\subsection*{D. Summary of design corollaries}
Taken together, the three design corollaries are purely kinematic consequences of the EPIC ledger and the homogeneities of transport and structural pricing. A uniform dilation enforces a \emph{universal} power split \(nP_{\rm flow}=mP_{\rm struct}\) [Eq.~\eqref{eq:equipartition}], so the fractions \(P_{\rm flow}/(P_{\rm flow}+P_{\rm struct})\) and \(P_{\rm struct}/(P_{\rm flow}+P_{\rm struct})\) depend only on \((m,n)\), independent of geometry or load. Eliminating radii at stationarity yields a flux–only edge cost \(\mathcal{C}^\star(Q;x)\propto\kappa_{\mathrm{hyd}}(x)^{\frac{m}{m+n}}[\varepsilon_b(x)c(x)]^{\frac{n}{m+n}}Q^{\gamma}\) with \(\gamma=2m/(m{+}n)\in(0,2)\) [Eq.~\eqref{eq:tariff_index}]: in the relevant regime \(m<n\), \(0<\gamma<1\) makes the cost strictly concave in \(Q\), so mergers are energetically preferred and the depth of tree–like hierarchies shifts systematically with \((m,n)\). Finally, the routing index is the \emph{dimensionless} composite field
\(\mathfrak n(x)=\big[\kappa_{\mathrm{hyd}}(x)/\kappa_{0}\big]^{\frac{m}{m+n}}
\big[\varepsilon_b(x)c(x)/(\varepsilon_{b0}c_{0})\big]^{\frac{n}{m+n}}\)
(reducing to \([\varepsilon_b(x)c(x)/(\varepsilon_{b0}c_{0})]^{\frac{n}{m+n}}\) when \(\kappa_{\mathrm{hyd}}\) is spatially uniform).
This converts path optimization into an “optical path length’’ problem [Eq.~\eqref{eq:EPIC_optical_length}], so spatial heterogeneity in the composite field induces Snell–type refraction \(\mathfrak n_1\sin\theta_1=\mathfrak n_2\sin\theta_2\) at sharp interfaces [Eq.~\eqref{eq:Snell}] and smooth bending toward cheaper regions in graded media. In this compact form, equipartition predicts how power budgets partition between pumping and upkeep; concavity explains why branching architectures dominate and how their depth responds to changes in upkeep scaling or rheology; and the routing index links observed conduit paths in heterogeneous tissues or engineered phantoms to an underlying price field in J/bit, in line with optimal–network perspectives such as Jiang and Li’s PRE review \cite{JiangLi2019PRE} and the PRL analysis of global optimization versus local adaptation by Ronellenfitsch and Katifori \cite{Ronellenfitsch2016PRL}.

\section{Conclusion}
\emph{EPIC changes the accounting, not the physics.}  The transport laws remain classical; what changes is the currency in which structure is priced. Auditing a node’s control layer in absolute units (J/bit) converts entropy production into a unitful budget for maintaining reliable distinctions, encoded in the node-level ledger of Eq.~\eqref{eq:EPIC_additive_node} with \(\Lambda\) and \(\varepsilon_b\) defined in Eqs.~\eqref{eq:calib_eb}–\eqref{eq:zeta_def}. 

From this priced extremum emerge three geometric outputs. The first is a weighted Murray family: the stationarity condition equalizes the branchwise quantity defined in Eq.~\eqref{eq:node_invariant}, which in turn yields the flow–radius scaling with exponent \(\alpha\) in Eq.~\eqref{eq:alpha} and the associated tariff–weighted Murray relation; in the Poiseuille case this recovers the classical cubic law for volume-priced upkeep together with its subcubic analogue when wall burden dominates, while for general hydraulic exponent \(n\) or a radius-dependent viscosity \(\mu(r)\) the effective slope is shifted as in Eq.~\eqref{eq:alpha_eff_exact_repeat}. A second output is a tariff-weighted angle law: translating the junction removes material prefactors and fixes the branch directions through the vector closure of Eq.~\eqref{eq:angle_vector}, predicting the near-symmetric daughter openings quoted above for homogeneous tariffs and providing a distinctive mixed-tariff signature when surface and volume contributions are blended. The third is a growth-rate bound: at any audit, the rate of creating stably maintained, reliably decodable distinctions is constrained by the two-faced growth-cone of Eq.~\eqref{eq:EPIC_cone_node}, so the achievable \(\dot I\) is set by the smaller of an interior bit budget purchasable from dissipation and a boundary budget determined by the reliable throughput across the causal surface.

Beyond these node laws, three design rules summarize how morphology responds to prices. Homogeneity alone enforces an equipartition identity \(nP_{\rm flow}=mP_{\rm struct}\), so the optimal power split between pumping and upkeep is universal; radii rescale as \(r\propto (\Lambda c)^{-1/(m+n)}\) at fixed demands, and the tariff–weighted angle law is unchanged. Eliminating \(r\) along an edge then yields an effective per–length cost \(\propto [\varepsilon_b c]^{\frac{n}{m+n}}\,Q^{\gamma}\) with \(\gamma=\tfrac{2m}{m+n}\); in the common \(m<n\) regime (e.g., Poiseuille with \(m{=}1\) or \(2\)) this exponent satisfies \(\gamma<1\), so the cost is strictly concave in \(Q\), mergers are energetically favored over isolated paths, and the branching depth shifts in a controlled way with upkeep mix \((m)\) and rheology \((n)\). The same elimination, at fixed \(Q\), defines the EPIC routing index \(\mathfrak n(x)\) of Eq.~\eqref{eq:EPIC_index_def}, a dimensionless composite price field that reduces to \([\varepsilon_b(x)c(x)/(\varepsilon_{b0}c_{0})]^{\frac{n}{m+n}}\) when \(\kappa_{\mathrm{hyd}}\) is spatially uniform. Optimal routes minimize the “optical path length’’ \(\int \mathfrak n(x)\,ds\) and obey a Snell–like refraction law \(\mathfrak n_1\sin\theta_1=\mathfrak n_2\sin\theta_2\), bending toward cheaper–tariff corridors.

All of these statements are explicitly unit bearing (J/bit, W, m) and admit preregistered tests: weighted–versus–unweighted Murray residuals; angle predictions and vector–closure with held–out directions; near–symmetric angle drift under controlled surface/volume perturbations; and refraction across engineered tariff jumps. The priorities are clear: (1) audit \(\varepsilon_b\) and \(c\) with uncertainties; (2) probe the surface–volume mix to move \(\alpha\) and the symmetric angle in the predicted directions; (3) map \(\mathfrak n(x)\) to forecast routing in heterogeneous media; and (4) identify when the boundary face, rather than the interior budget, limits \(\dot I\). In this sense EPIC supplies a calibrated ledger (J/bit) that prices structure, restores Murray’s law as a special case, fixes angles, bounds growth, and furnishes simple design rules (equipartition, concavity, refraction) that render morphology calculable and testable in the laboratory.

\section{Notation and units}\label{sec:notation}
\begingroup
\footnotesize
\begin{description}[leftmargin=2.4cm,labelsep=0.6cm,align=left,font=\normalfont,style=unboxed,wide=0pt]

  \item[\(Q\)] Volumetric flow rate \textemdash{} segment–level demand (fixed daughters) \textemdash{} \si{\metre\cubed\per\second}.
  \item[\(r,\,\ell\)] Segment radius and length \textemdash{} long–tube (Poiseuille) assumption \textemdash{} \si{\metre}.
  \item[\(\mu\)] Dynamic viscosity \textemdash{} \si{\pascal\second}; may depend on \(r\) in microvasculature.
  \item[\(\Delta p\)] Pressure drop per segment \textemdash{} \si{\pascal}.
  \item[\(P_{\rm flow}\)] Hydraulic (dissipated) power \textemdash{} \(\sum_i Q_i\,\Delta p_i\) \textemdash{} \si{\watt}.
  \item[\(\sigma_s\)] Entropy–production density \textemdash{} local, non–negative \textemdash{} \si{\watt\per\kelvin\per\metre\cubed}.
  \item[\(T\)] Temperature \textemdash{} node–isothermal audit temperature \textemdash{} \si{\kelvin}.

  \item[\(\Psi_b\)] Information throughput density \textemdash{} \(\Psi_b=\sigma_s T/\varepsilon_b\) \textemdash{} \si{\bit\per\metre\cubed\per\second}.
  \item[\(\varepsilon_b\)] Effective bit energy (tariff) \textemdash{} \(\varepsilon_b=\zeta k_B T\ln 2\) \textemdash{} \si{\joule\per\bit}.
  \item[\(\eta_b\)] Structural information efficiency \textemdash{} \(\eta_b=\varepsilon_b^{-1}\) \textemdash{} \si{\bit\per\joule}.
  \item[\(\zeta\)] Overhead factor \textemdash{} dimensionless, \(\zeta\ge 1\) (speed, reliability, redundancy, control).
  \item[\(\beta\)] Bit–rate density \textemdash{} audit–level rate of reliable operations \textemdash{} \si{\bit\per\metre\cubed\per\second}.

  \item[\(\mathcal B\)] Structural bit–rate (node) \textemdash{} \(\mathcal B=\sum_i c_i\,\ell_i\,r_i^{m}\) \textemdash{} \si{\bit\per\second}.
  \item[\(c_i\)] Tariff coefficient on branch \(i\) \textemdash{} in \(\mathcal B_i=c_i\,\ell_i\,r_i^{m}\) \textemdash{} \([\mathrm{bit}\,\mathrm{s}^{-1}\,\mathrm{m}^{-(m+1)}]\).
  \item[\(m\)] Upkeep exponent \textemdash{} dimensionless; \(m=2\) (volume–priced), \(m=1\) (surface–priced).
  \item[\(n\)] Hydraulic exponent \textemdash{} dimensionless; Poiseuille \(n=4\) (other \(n\) for slip/non–Newtonian).
  \item[\(\Lambda\)] Priced–bit multiplier \textemdash{} \(\Lambda=\lambda\,\varepsilon_b\) \textemdash{} \si{\joule\per\bit}.
  \item[\(\mathcal{K}\)] Stationarity invariant \textemdash{} \(Q_i^2/(c_i r_i^{m+4})\) \textemdash{} \si{\metre^{3}\per\second\per\bit}.

  \item[\(\tau_w\)] Wall shear stress \textemdash{} \(4\mu Q/(\pi r^3)\) (laminar) \textemdash{} \si{\pascal}.
  \item[\(\dot S_{\mathrm{tot}}\)] Total entropy–production rate \textemdash{} \(\int_V \sigma_s\,dV\) \textemdash{} \si{\watt\per\kelvin}.
  \item[\(P_{\rm diss}\)] Dissipated power (domain) \textemdash{} \(\int_V \sigma_s T\,dV\) \textemdash{} \si{\watt}.
  \item[\(I,\,\dot I\)] Stock of reliably stored distinctions and its creation rate \textemdash{} \si{\bit}, \si{\bit\per\second}.

  \item[\(\theta\)] Junction opening angle \textemdash{} between branch axes \textemdash{} \si{\degree}.
  \item[\(\mathbf e_i\)] Unit direction vector along branch \(i\) \textemdash{} dimensionless.

  \item[\(\kappa_{\mathrm{hyd}}(x)\)]
    Hydraulic prefactor per unit length in \(\mathcal C(Q,r;x)=\kappa_{\mathrm{hyd}}(x)Q^{2}r^{-n}+\Lambda c(x)r^{m}\), chosen so that \(\kappa_{\mathrm{hyd}} Q^{2} r^{-n}\) has units \si{\watt\per\metre}.
    \emph{Units:} \([\,\kappa_{\mathrm{hyd}}\,]=\si{kg\,m^{\,n-5}\,s^{-1}}\).
    For Poiseuille (\(n{=}4\)), \([\,\kappa_{\mathrm{hyd}}\,]=\si{\pascal\second}\) and \(\kappa_{\mathrm{hyd}}=8\mu/\pi\).

  \item[\(\kappa_{0}\)] Reference hydraulic prefactor (per unit length) used to nondimensionalize \(\mathfrak n\); same units as \(\kappa_{\mathrm{hyd}}\).
  \item[\(\varepsilon_{b0}\)] Reference bit tariff used to nondimensionalize \(\mathfrak n\); \([\varepsilon_{b0}]=\si{J\,bit^{-1}}\).
  \item[\(c_{0}\)] Reference structural coefficient used to nondimensionalize \(\mathfrak n\); \([c_{0}]=\si{bit\,s^{-1}\,m^{-(m+1)}}\).
  \item[\(\mathfrak n(x)\)] EPIC routing (refractive) index — \emph{dimensionless} composite price field; see Eq.~\eqref{eq:EPIC_index_def}. Optimal routes minimize \(\int \mathfrak n(x)\,ds\).

\end{description}
\vspace{2pt}
{\small\emph{Note}: 1 bit \(=\ln 2\) nats (equivalently, 1 nat \(=1/\ln 2\) bits); convert via \(\mathrm{nats}=\mathrm{bits}\times \ln 2\).}
\endgroup

\section{Methods: calibration and reliability}
\phantomsection\label{sec:calibration}

This section specifies how dissipated power is translated into an information–throughput budget, how the effective bit energy \(\varepsilon_b\) is operationally defined, and how reliability and overheads are reported. The goal is to make the EPIC tariff \(\varepsilon_b\) concrete enough that any platform with metered control-layer power and a measurable reliable bit-rate can, in principle, reproduce the same calibration. A short worked example at the end applies the protocol to a single Y-junction in the Poiseuille regime. When a control‑layer audit is unavailable, relative tariffs at a junction can be inferred purely from geometry; see Appendix~\ref{app:tomography}.

The audit begins by fixing a time window of duration \(\tau>0\) over which both the control layer and the transport remain in a statistically steady operating state, and by choosing a node-isothermal audit temperature \(T\). Over this window, let \(\langle P_{\rm ctrl}\rangle_\tau\) denote the \emph{baseline-subtracted} mean power attributable specifically to sensing, actuation, monitoring, and housekeeping—i.e., the part of the power budget devoted to maintaining reliable distinctions rather than to the bulk transport itself. Let \(\langle\dot{\mathcal I}_{\rm rel}\rangle_\tau\) denote the mean rate, over the same window, at which reliably decodable bits are produced at the chosen error target. The operational energetic tariff per reliable bit is then defined as
\begin{align}
\label{eq:calib_eb}
\varepsilon_b
&=\frac{\langle P_{\rm ctrl}\rangle_\tau}{\langle\dot{\mathcal I}_{\rm rel}\rangle_\tau}
\qquad [\si{J\,bit^{-1}}],\\[2pt]
\label{eq:zeta_def}
\zeta
&\equiv \frac{\varepsilon_b}{k_B T\ln 2}\ \ge\ 1 ,
\end{align}
so that \(\zeta\) measures the overhead relative to the Landauer limit \(k_B T\ln 2\) per erased bit. In these units, \(\varepsilon_b\) converts watts into reliably decodable bits per second via its reciprocal \(\eta_b\equiv\varepsilon_b^{-1}\) [\si{bit\,J^{-1}}]: each watt of control-layer power sustains at most \(\eta_b\) reliable bits per second at the audited reliability.

When reliability is specified by a per-symbol (bit) error probability \(p_e\in[0,\tfrac12)\), the fraction of each raw bit that can be regarded as reliably decodable is \(1-\Hb(p_e)/\ln 2\), where
\[
\Hb(p)=-\,p\ln p-(1-p)\ln(1-p)
\]
is the binary entropy in nats. In other words, only the “distance” from full uncertainty contributes to reliably decodable information. The corresponding multiplicative factor that maps the raw bit-rate to a reliability-normalized bit-rate is
\begin{equation}
\label{eq:zeta_rel}
\zeta_{\rm rel}
=\bigl[\,1-\Hb(p_e)/\ln 2\,\bigr]^{-1}\ \ge\ 1 ,
\end{equation}
so that tightening the error target (\(p_e \downarrow 0\)) decreases \(\zeta_{\rm rel}\) toward \(1\) (not increases), while relaxing reliability (\(p_e \uparrow \tfrac12\)) makes \(\zeta_{\rm rel}\) grow without bound; accordingly this factor decreases or increases \(\varepsilon_b\), respectively.

The overhead \(\zeta\) is dimensionless and admits a natural factorization into conceptually distinct contributions,
\begin{equation}
\label{eq:zeta_factorization}
\zeta \;=\; \zeta_{\rm speed}\,\zeta_{\rm rel}\,\zeta_{\rm red}\,\zeta_{\rm ctrl},
\qquad
\zeta_\bullet \ge 1 \, ,
\end{equation}
where \(\zeta_{\rm speed}\) accounts for finite-time or throughput premiums (operating the control layer faster than its quasi-static optimum), \(\zeta_{\rm rel}\) is the reliability normalization from Eq.~\eqref{eq:zeta_rel}, \(\zeta_{\rm red}\) encodes redundancy and proofreading overheads (repeated reads, majority voting, error-correcting codes), and \(\zeta_{\rm ctrl}\) aggregates standing control and housekeeping costs that are not easily attributed to individual bit operations. Taking logarithms gives a simple bookkeeping identity,
\[
\ln\zeta
= \ln\zeta_{\rm speed} + \ln\zeta_{\rm rel}
+ \ln\zeta_{\rm red} + \ln\zeta_{\rm ctrl},
\]
which makes it transparent how improvements in device physics, coding, or control can be reported as explicit reductions in one or more of these factors.

Once \(\varepsilon_b\) is fixed, local entropy production is converted into an information-throughput density by pricing dissipation at the audited tariff. The power-priced information throughput density is defined as
\begin{equation}
\label{eq:Psi_def_repeat}
\Psi_b(x,t)
\;\equiv\;
\frac{\sigma_s(x,t)\,T(x,t)}{\varepsilon_b}
\qquad [\si{\bit\per\metre\cubed\per\second}] ,
\end{equation}
where \(\sigma_s(x,t)\) [\si{W\,K^{-1}\,m^{-3}}] is the local entropy-production density and \(T(x,t)\) is the local audit temperature. In steady, isothermal domains this density integrates to a global interior bit budget,
\begin{equation}
\label{eq:Psi_budget}
\int_V \Psi_b\,dV
\;=\;\frac{\int_V \sigma_s T\,dV}{\varepsilon_b}
\;=\;\frac{P_{\rm diss}}{\varepsilon_b}
\;=\;\frac{P_{\rm flow}}{\varepsilon_b}\,,
\end{equation}
where \(P_{\rm diss}=\int_V \sigma_s T\,dV\) is the total dissipated power and, in the present hydraulic setting, \(P_{\rm flow}=\sum_i Q_i\Delta p_i\) is the familiar pumping power. Equation~\eqref{eq:Psi_budget} expresses the core exchange rate: at the audited tariff, each watt of dissipation can support at most \(\eta_b=\varepsilon_b^{-1}\) reliable bits per second. Because \(\varepsilon_b=\zeta k_B T\ln 2\), the explicit temperature cancels in \(\Psi_b=\sigma_s/(\zeta k_B\ln 2)\); at fixed audit, only the dissipation density \(\sigma_s\) and the overhead \(\zeta\) determine the interior information budget. In heterogeneous settings with spatially varying \(\varepsilon_b(x)\) the relevant quantity is \(\int_V \sigma_s(x) T(x)/\varepsilon_b(x)\,dV\), i.e., dissipation weighted by the local price field.

To illustrate this calibration in concrete SI units, consider a single Y-junction operating in the Hagen–Poiseuille regime. The purpose of the example is not to model any specific biological vessel, but to display how a standard Poiseuille calculation for pumping power translates directly into a reliability-normalized bit-throughput once \(\varepsilon_b\) is specified. Assume steady, fully developed, laminar, Newtonian flow in long circular tubes as in standard microfluidics references~\cite{Bruus2008,Kirby2010}. Take
\begin{gather*}
\mu   = \qty{3.0}{mPa.s},\qquad
\ell_i = \qty{1.0}{mm},\\[2pt]
r_1 = \qty{12}{\micro m},\quad
r_2 = \qty{10}{\micro m},\quad
r_0 = \qty{15}{\micro m},\\[2pt]
Q_1 = \qty{5}{pL/s},\quad
Q_2 = \qty{3}{pL/s},\quad
Q_0 = Q_1 + Q_2 = \qty{8}{pL/s}.
\end{gather*}
For each segment \(i\), the Hagen–Poiseuille relations give
\begin{align}
\Delta p_i
&= \frac{8\mu\,\ell_i}{\pi}\,\frac{Q_i}{r_i^4},
\\[2pt]
P_{{\rm flow},i}
&= Q_i\,\Delta p_i
 = \frac{8\mu\,\ell_i}{\pi}\,\frac{Q_i^{2}}{r_i^{4}}.
\end{align}
With the numerical values above,
\begin{gather}
\Delta p_0 = \qty{1.21}{kPa},\quad
\Delta p_1 = \qty{1.84}{kPa},\quad
\Delta p_2 = \qty{2.29}{kPa},\\[2pt]
P_0 = \qty{9.66}{nW},\quad
P_1 = \qty{9.21}{nW},\quad
P_2 = \qty{6.88}{nW},\\[2pt]
P_{\rm flow} = \sum_i P_i = \qty{25.74}{nW}.
\end{gather}
A quick dimensional check,
\begin{align*}
[P]
&= [\mu][\ell]\,\frac{Q^2}{r^4}
 = (\si{Pa.s})(\si{m})\,\frac{\si{m^6.s^{-2}}}{\si{m^4}} \\
&= \si{N.m.s^{-1}} = \si{W},
\end{align*}
confirms that the units are consistent.

At the audit temperature \(T=\qty{310}{K}\), Landauer’s bound is \(k_B T\ln 2=\qty{2.967e-21}{J/bit}\). If one reliable bit is priced at \(\varepsilon_b=\zeta\,k_BT\ln 2\), the junction’s interior information budget in steady state is
\[
\frac{P_{\rm flow}}{\varepsilon_b}
=\frac{\qty{2.574e-8}{W}}{\zeta\,\qty{2.967e-21}{J/bit}}
=\frac{8.68\times 10^{12}}{\zeta}\ \si{bit/s}.
\]
Reporting \(\zeta\) as a bracketed range makes the dependence transparent: for \(\zeta\in[10^3,10^6]\),
\[
\frac{P_{\rm flow}}{\varepsilon_b}
\in \bigl[\,8.68\times 10^{6},\,8.68\times 10^{9}\,\bigr]\ \si{bit/s}.
\]

The hydraulic regime assumed in this example is internally consistent. The mean speeds \(u_i=Q_i/(\pi r_i^2)\) are approximately \(\qty{1.13}{cm/s}\), \(\qty{1.10}{cm/s}\), and \(\qty{0.95}{cm/s}\) for \(i=0,1,2\), respectively. With \(\rho\approx\qty{1000}{kg/m^3}\), the Reynolds numbers \(\mathrm{Re}_i=\rho u_i(2r_i)/\mu\) are \((0.113,\,0.088,\,0.064)\ll 1\), confirming creeping-to-laminar flow~\cite{Bruus2008,Kirby2010}. Hydrodynamic entrance lengths \(L_e\sim 0.05\,\mathrm{Re}\,D\) are sub‑micron~\cite{Bruus2008,Kirby2010}, so over \(\qty{1}{mm}\) segments the velocity profiles are fully developed. The corresponding wall shear stresses
\(\tau_{w,i}=4\mu Q_i/(\pi r_i^3)\) take values \((\qty{9.1}{Pa},\,\qty{11.0}{Pa},\,\qty{11.5}{Pa})\), consistent in order of magnitude with in vivo arteriolar ranges (typically \(\sim\!1\text{–}5\,\si{Pa}\) depending on vessel class and conditions~\cite{Reneman2008WSS}). Because \(P_{{\rm flow},i}\propto Q_i^2 r_i^{-4}\), small fractional errors in radius dominate the numerical uncertainty (for example, a \(\pm 10\%\) error in \(r_i\) shifts \(P_{{\rm flow},i}\) by roughly \(\mp 40\%\)). None of these details alter the underlying accounting: once \(P_{\rm flow}\) and \(\varepsilon_b\) are independently audited, the junction’s interior information budget is \(\dot I_{\rm int}=\eta_b P_{\rm flow}\) with \(\eta_b=\varepsilon_b^{-1}\). In other words, watts in are read directly as reliably decodable \si{bit/s} out at the stated tariff and reliability, completing the link between hydraulic dissipation and the EPIC currency.

\appendix

\section{Proof of the growth-cone bound}
\label{sec:cone_proof}
This section states explicit premises and derives the instantaneous two–face bound on the rate of creating stably maintained, reliably decodable distinctions inside a space–time control volume.

\paragraph*{Assumptions (A1–A4).}
Let \(V\) be a domain with (time-dependent) boundary \(\Sigma_t\), and let \(I(t)=\int_V \iota(x,t)\,dV\) count \(\epsilon\)–reliable, stably stored distinctions at the audit’s resolution.
\begin{enumerate}[label=(A\arabic*),leftmargin=*,itemsep=2pt,topsep=2pt]
\item \emph{Audit \& tariff.} \(\varepsilon_b(x,t)=\zeta(x,t)\,k_B T(x,t)\ln 2\) with \(\zeta\!\ge\!1\) measured as in \eqref{eq:calib_eb}–\eqref{eq:zeta_def}.
\item \emph{Information share of dissipation.} There exists a measurable density \(\dot{\mathcal I}_v(x,t)\) \([\si{bit\,s^{-1}\,m^{-3}}]\) of reliably executed operations such that the information–processing share of entropy production covers the tariff,
\begin{equation}
\label{eq:M1}
\sigma_s^{\rm(info)}(x,t)\ \ge\ \frac{\varepsilon_b(x,t)}{T(x,t)}\,\dot{\mathcal I}_v(x,t).
\end{equation}
\item \emph{Reliable boundary communication.} Across \(\Sigma_t\), physical channels operated at blocklength \(N\) and error \(\epsilon\) deliver reliable rate density \(\mathcal C_\epsilon(N;x,t)\) \([\si{bit\,s^{-1}\,m^{-2}}]\).
\item \emph{Stock of distinctions.} \(I(t)\) increases only when reliably decodable bits are \emph{written} to storage by interior operations. Any boundary inflow must be decoded and written, so the net rate is simultaneously limited by interior write capacity and by boundary throughput:
\[
\dot I(t)\ \le\ \min\!\left\{\ \int_V \dot{\mathcal I}_v(x,t)\,dV\ ,\ \int_{\Sigma_t}\! \mathcal C_\epsilon(N;x,t)\,dA\ \right\}.
\]
\end{enumerate}

\paragraph*{Lemma 1 (interior cut).}
Integrating \eqref{eq:M1} over \(V\) and using \(\int_V\sigma_s^{\rm(info)}\le \int_V \sigma_s\) yields
\begin{equation}
\label{eq:M2}
\int_V \dot{\mathcal I}_v\,dV \ \le\ \int_V \frac{\sigma_s T}{\varepsilon_b}\,dV \ =\ \int_V \Psi_b\,dV.
\end{equation}
Since the rate of growth of the stock cannot exceed the rate of reliable operations,
\begin{equation}
\label{eq:M3}
\dot I(t)\ \le\ \int_V \Psi_b(x,t)\,dV.
\end{equation}

\paragraph*{Lemma 2 (boundary cut).}
Let \(M(t,\Delta t)\) be the reliably decodable message entering through \(\Sigma_t\) during \([t,t+\Delta t]\). By data processing and the reliability specification,
\begin{equation}
\label{eq:M4}
\mathbb{E}[\Delta I]\ \le\ \Delta t\int_{\Sigma_t}\! \mathcal C_\epsilon(N;x,t)\,dA.
\end{equation}
Dividing by \(\Delta t\) and sending \(\Delta t\downarrow 0\) gives
\begin{equation}
\label{eq:M5}
\dot I(t)\ \le\ \int_{\Sigma_t}\! \mathcal C_\epsilon(N;x,t)\,dA.
\end{equation}

\paragraph*{Theorem (Growth–cone).}
Combining \eqref{eq:M3} and \eqref{eq:M5} yields the instantaneous bound
\begin{equation}
\frac{d}{dt}\!\int_V \iota\,dV
\ \le\
\min\!\left\{\ \int_V \Psi_b(x,t)\,dV\ ,\ \int_{\Sigma_t}\! \mathcal C_\epsilon(N;x,t)\,dA\ \right\}.
\end{equation}
In a steady, isothermal node, \(\int_V \Psi_b\,dV=P_{\rm diss}/\varepsilon_b=P_{\rm flow}/\varepsilon_b\).

\paragraph*{Finite-blocklength approximation (for \(\mathcal C_\epsilon\)).}
With channel density \(\rho_{\rm ch}(x,t)\), symbol rate \(W\), asymptotic capacity \(C\) (bits/use), and dispersion \(V\), the normal approximation gives
\[
\mathcal C_\epsilon(N;x,t)
=\rho_{\rm ch}(x,t)\,W(x,t)\,
\Big[\,C-\sqrt{V/N}\,\Phi^{-1}(\epsilon)\,\Big]_+,
\]
with \(\Phi^{-1}\) the Gaussian quantile and \([a]_+\equiv\max\{a,0\}\). Equality in the growth–cone requires: (i) the nonlimiting face is slack (e.g., boundary not rate-limiting when the interior face is active), (ii) the limiting face operates reversibly at the audit’s reliability, and (iii) no additional finite–time or idle-duty overheads beyond the audited \(\zeta\) erode the budget.

\paragraph*{Worked example: growth–cone bound for a junction-sized domain.}
To make the bound concrete, consider a control volume \(V\) that encloses the single Y-junction used in the node audit above. In that example the total hydraulic power was
\(P_{\rm flow}=\qty{25.74}{nW}\), so at an audited tariff \(\varepsilon_b=\zeta k_B T\ln 2\) the interior bit budget is
\[
\dot I_{\rm int}
\equiv \int_V \Psi_b\,dV
= \frac{P_{\rm flow}}{\varepsilon_b}
= \frac{8.68\times 10^{12}}{\zeta}\ \si{bit/s}.
\]
For definiteness, take \(\zeta=10^{4}\), which corresponds to an energetic cost of \(\varepsilon_b\approx\qty{29.7}{fJ/bit}\) at \(T=\qty{310}{K}\); then \(\dot I_{\rm int}\approx \qty{8.68e8}{bit/s}\).

Now let \(\Sigma_t\) be a notional “instrumented” surface of area \(A=\qty{1.0}{mm^2}=\qty{1.0e-6}{m^2}\) carrying a uniform density \(\rho_{\rm ch}\) of independent physical channels (e.g., electrode lines or optical links) that report from the domain to the outside world. For clarity, assume that over \(\Sigma_t\) the channel parameters in the finite-blocklength approximation \cite{CoverThomas2006,Polyanskiy2010} are constant:
\[
\mathcal C_\epsilon(N;x,t)
=\rho_{\rm ch}\,W\,
\Big[\,C-\sqrt{V/N}\,\Phi^{-1}(\epsilon)\,\Big]_+,
\]
so that
\[
\int_{\Sigma_t}\! \mathcal C_\epsilon\,dA
= A\,\rho_{\rm ch}\,W\,
\Big[\,C-\sqrt{V/N}\,\Phi^{-1}(\epsilon)\,\Big]_+.
\]
As a concrete choice, let
\begin{align}
\rho_{\rm ch} &= 10^{8}\ \si{m^{-2}}, & 
W &= 10^{4}\ \si{s^{-1}}, \\
C &= 1\ \si{bit}\,\text{per use},   & 
V &\approx 1\ \si{bit^{2}}\,\text{per use}, \\
N &= 10^{4},            & 
\epsilon &= 10^{-6}.
\end{align}
The normal approximation then gives a penalty term
\(\sqrt{V/N}\,\Phi^{-1}(\epsilon)\approx 0.01\times 4.75\approx 0.048\), so the effective per-use rate is \(\approx 0.95\ \si{bit/use}\). The boundary face becomes
\begin{align*}
\int_{\Sigma_t}\! \mathcal C_\epsilon\,dA
&\approx A\,\rho_{\rm ch}\,W\times 0.95\\
&= \qty{1.0e-6}{m^2}\times 10^{8}\ \si{m^{-2}}
   \times 10^{4}\ \si{use/s}\times 0.95\\
&\approx 9.5\times 10^{5}\ \si{bit/s}.
\end{align*}

The growth–cone bound
\[
\dot I(t)
\;\le\;
\min\!\left\{
\int_V \Psi_b\,dV\ ,\
\int_{\Sigma_t}\! \mathcal C_\epsilon\,dA
\right\}
\]
therefore reads, for these numbers,
\[
\dot I(t)
\;\le\;
\min\!\left\{
\,8.68\times 10^{8}\ \si{bit/s}\ ,\
9.5\times 10^{5}\ \si{bit/s}\,
\right\}
\approx 9.5\times 10^{5}\ \si{bit/s}.
\]
In this regime the \emph{boundary} is rate-limiting: the domain could, in principle, fund \(\mathcal O(10^9)\) reliable \si{bit/s} from its interior dissipation at the audited tariff, but the physical channels on \(\Sigma_t\) can convey only \(\sim 10^{6}\ \si{bit/s}\). If, instead, the interior dissipation or \(\varepsilon_b^{-1}\) were reduced so that \(P_{\rm flow}/\varepsilon_b \ll 10^{6}\ \si{bit/s}\), the same inequality would become \emph{interior}-limited. This illustrates the intended use of the bound: whichever face (interior or boundary) yields the smaller budget at the audited \(\varepsilon_b\) limits the instantaneous growth of stably stored, reliably decodable distinctions.

\section{Rheology details and general \texorpdfstring{$n$}{n}}
\label{app:rheology}

This appendix gives a compact, self‑contained derivation of how (i) a radius‑dependent viscosity \(\mu=\mu(r)\) and (ii) a generalized hydraulic exponent \(n\) in the per–length dissipation \(P_{\rm flow}\propto Q^2 r^{-n}\) enter the node optimality conditions. Throughout, results are derived for a single junction with fixed daughter demands \((Q_1,Q_2)\) and \(Q_0=Q_1+Q_2\).

The derivation begins with the exact stationarity conditions when \(\mu\) varies
with radius. At fixed \(Q_i\) the node objective reads
\[
\mathcal L(\mathbf r)
=\sum_{i=0}^2 \frac{8\ell_i}{\pi}\,\mu(r_i)\,\frac{Q_i^2}{r_i^4}
\;+\;\Lambda\sum_{i=0}^2 c_i\,\ell_i\,r_i^{m}.
\]
Setting \(\partial \mathcal L/\partial r_i=0\) and cancelling \(\ell_i>0\) yields
\[
\frac{8}{\pi}Q_i^2\!\left[\mu'(r_i)\,r_i^{-4}-4\mu(r_i)\,r_i^{-5}\right]
+\Lambda c_i m r_i^{m-1}=0,
\]
which rearranges to the exact branchwise invariant
\begin{equation}
\label{eq:EPIC_invariant_mu}
\frac{Q_i^2}{c_i\,r_i^{m+4}}
=\frac{\pi\,m\,\Lambda}{8\,\big[\,4\,\mu(r_i)-r_i\,\mu'(r_i)\,\big]}
\ \equiv\ \mathcal{K}_{\rm eff}(r_i).
\end{equation}
When \(\mu'(r)\equiv 0\) this reduces to the constant–viscosity result; in general the rheology enters only through
\(D(r)\equiv 4\mu(r)-r\mu'(r)\).

Taking a logarithmic derivative of \eqref{eq:EPIC_invariant_mu} with respect to \(\ln r\) gives the exact local flow–radius slope
\begin{equation}
\label{eq:alpha_eff_exact}
\alpha_{\rm eff}(r)\ \equiv\ \frac{d\ln Q}{d\ln r}
=\frac{1}{2}\left[(m{+}4)-\frac{d}{d\ln r}\ln D(r)\right].
\end{equation}
Hence, if \(D(r)\) increases with \(r\) then \(\alpha_{\rm eff}<\tfrac{m+4}{2}\); if \(D(r)\) decreases, \(\alpha_{\rm eff}>\tfrac{m+4}{2}\). In the slow‑variation regime write \(D(r)=4\mu(r)[1-\xi(r)]\) with \(|\xi|\ll 1\) and slowly varying \(\xi\). Using \(\frac{d}{d\ln r}\ln D=\frac{d\ln\mu}{d\ln r}+\mathcal O(\xi,\tfrac{d\xi}{d\ln r})\) one obtains
\begin{equation}
\label{eq:alpha_eff_slowmu}
\alpha_{\rm eff}(r)
=\frac{m+4}{2}-\frac{1}{2}\frac{d\ln\mu}{d\ln r}
+\mathcal O\!\Big(\xi,\ \xi^2,\ \tfrac{d\xi}{d\ln r}\Big),
\end{equation}
so an increasing \(\mu(r)\) depresses the slope relative to the Poiseuille value \((m{+}4)/2\), while a decreasing \(\mu(r)\) elevates it.

A complementary, length‑free diagnostic follows by combining the laminar wall‑shear formula \(\tau_w=4\,\mu(r)\,Q/(\pi r^3)\) with \eqref{eq:EPIC_invariant_mu}. This reproduces the priced–shear invariant \(\mathscr S_i\) defined in Eq.~\eqref{eq:priced_shear_invariant_repeat}, which is the same for \(i=0,1,2\) and, under the present conventions, has units \(\sqrt{\si{J/bit}}\). For equal tariffs \(c_i=c\) and radius‑independent viscosity \(\mu'(r)\equiv 0\), Eq.~\eqref{eq:priced_shear_invariant_repeat} reduces to \(\tau_{w,i}\,r_i^{(2-m)/2}=\text{const}\); in particular, \(m=2\) recovers uniform wall shear, while \(m=1\) gives \(\tau_w r^{1/2}=\text{const}\). When \(\mu(r)\) varies, the multiplicative factor \(\sqrt{4\mu(r_i)-r_i\mu'(r_i)}/\mu(r_i)\) modulates this condition. Thus \(\mathscr S_i\) packages geometry \((r_i)\), tariffs \((c_i)\), and rheology \((\mu,\mu')\) into a single, segment‑length–independent stationarity check consistent with \eqref{eq:EPIC_invariant_mu} and \eqref{eq:alpha_eff_exact}.

The same variational structure also accommodates a generalized hydraulic exponent. Let the per–length dissipation scale as \(Q^2 r^{-n}\) (Poiseuille \(n=4\); slip or non‑Newtonian effects give \(n\neq 4\)). The KKT condition at fixed \(Q\) becomes
\[
\frac{\partial}{\partial r}\!\left(Q^2 r^{-n}\right)+\Lambda\,c\,m\,r^{m-1}=0,
\]
which yields the branch relation \(Q\propto \sqrt{c}\,r^{(m+n)/2}\), i.e.
\begin{equation}
\label{eq:alpha_general_n}
\alpha=\frac{m+n}{2}.
\end{equation}
Angles remain controlled solely by translation stationarity,
\(\sum_i c_i r_i^{m}\,\mathbf e_i=\mathbf 0\), independent of \(n\). Moreover, a uniform dilation \(r\mapsto(1{+}\epsilon)r\) at the interior optimum gives the homogeneity (equipartition) identity
\begin{equation}
\label{eq:equipartition_appendix}
n\,P_{\rm flow}\;=\;m\,P_{\rm struct},
\qquad P_{\rm struct}\equiv \Lambda\,\mathcal B,
\end{equation}
so the optimal power fractions are \(P_{\rm flow}/(P_{\rm flow}{+}P_{\rm struct})=m/(m{+}n)\) and \(P_{\rm struct}/(P_{\rm flow}{+}P_{\rm struct})=n/(m{+}n)\).

In summary, rheology enters the node problem through the single factor \(D(r)=4\mu-r\mu'\) in the exact invariant \eqref{eq:EPIC_invariant_mu}; this determines both the local flow–radius slope \eqref{eq:alpha_eff_exact}–\eqref{eq:alpha_eff_slowmu} and the priced–shear diagnostic via Eq.~\eqref{eq:priced_shear_invariant_repeat}. By contrast, the geometrical angle law and the homogeneity‑based equipartition \eqref{eq:equipartition_appendix} are insensitive to \(n\) and to \(\mu(r)\), reflecting the separation between tariff‑weighted geometry and transport rheology in the EPIC extremum.

\section{Geometry-only tariff tomography}
\label{app:tomography}
This appendix shows how to infer the \emph{relative} tariffs \(c_1/c_0\) and \(c_2/c_0\) from a single node when only \((r_0,r_1,r_2,\theta_{12})\) are measured and no control-layer audit is available. Throughout \(m>0\) is fixed, \(\alpha=\tfrac{m+4}{2}\), radii are strictly positive, and \(0<\theta_{12}<\pi\).

\paragraph*{Setup and normalization.}
Let
\[
k_1\equiv \frac{c_1}{c_0},\qquad
k_2\equiv \frac{c_2}{c_0},\qquad
x\equiv \sqrt{k_1}>0,\qquad
y\equiv \sqrt{k_2}>0 .
\]
Define the compact notation
\[
A=r_1^{\alpha},\quad B=r_2^{\alpha},\quad C=r_0^{\alpha},\qquad
W_1=r_1^{m},\quad W_2=r_2^{m}.
\]
The two node laws are
\begin{align}
\label{eq:tom_wM}
\text{(weighted Murray)}\qquad
&C=\;xA+yB,\\[2pt]
\label{eq:tom_cos}
\text{(cosine law)}\qquad
&\cos\theta_{12}
=\frac{r_0^{2m}-x^{4}W_1^{2}-y^{4}W_2^{2}}{2\,x^{2}y^{2}\,W_1W_2}\, .
\end{align}
Equation \eqref{eq:tom_wM} expresses \(y\) as an affine function of \(x\):
\begin{equation}
\label{eq:y_of_x}
y(x)=\frac{C-xA}{B}\,,\qquad 0<x<\frac{C}{A},
\end{equation}
where the open interval ensures \(x>0\) and \(y(x)>0\).

\paragraph*{Single–variable reduction.}
Substituting \eqref{eq:y_of_x} into \eqref{eq:tom_cos} yields a scalar equation for \(x\):
\begin{equation}
\label{eq:phi_def}
\begin{aligned}
\Phi(x)
&\equiv
\frac{r_0^{2m}-x^{4}W_1^{2}-y(x)^{4}W_2^{2}}
     {2\,x^{2}\,y(x)^{2}\,W_1W_2}
-\cos\theta_{12}
=0,\\
&\text{with } x\in\Bigl(0,\tfrac{C}{A}\Bigr),\quad y(x)=\frac{C-xA}{B}.
\end{aligned}
\end{equation}
Clearing denominators shows that \(\Phi\) is a real quartic rational function of \(x\); it is smooth on the admissible interval.

\paragraph*{Compatibility, bracketing, and solution structure.}
For the common morphological case \(r_0>\max\{r_1,r_2\}\) (often observed empirically; it is not entailed by \eqref{eq:tom_wM}–\eqref{eq:tom_cos}), the endpoint limits are
\[
\lim_{x\downarrow 0}\Phi(x)=-\infty,\qquad
\lim_{x\uparrow C/A}\Phi(x)=-\infty,
\]
because the denominator \(2x^{2}y(x)^{2}W_1W_2\!\downarrow 0^+\) while the numerator remains finite and strictly negative:
\[
\lim_{x\downarrow 0}\bigl[r_0^{2m}-x^{4}W_1^2-y(x)^{4}W_2^2\bigr]
=r_0^{2m}\!\left[1-\biggl(\frac{r_0}{r_2}\biggr)^{\!8}\right]\!<0,
\]
and symmetrically at \(x\uparrow C/A\) with \(r_1\) in place of \(r_2\). Hence the reduced cosine
\[
f(x)\;\equiv\;
\frac{r_0^{2m}-x^{4}W_1^{2}-y(x)^{4}W_2^{2}}
     {2\,x^{2}\,y(x)^{2}\,W_1W_2}
\]
satisfies \(f(x)\to-\infty\) at both ends of the interval and is continuous on \((0,C/A)\); therefore it attains at least one interior maximum \(f_{\max}=f(x^\ast)\).

\emph{Compatibility condition.} A solution \((x,y)\) to \eqref{eq:tom_wM}–\eqref{eq:tom_cos} exists \emph{iff}
\begin{equation}
\label{eq:compat}
\cos\theta_{12}\ \le\ f_{\max}\;\equiv\;\sup_{x\in(0,C/A)} f(x).
\end{equation}
Equivalently, \(\Phi(x)=f(x)-\cos\theta_{12}\) has a zero if and only if \(\sup_{x}\Phi(x)\ge 0\). When \eqref{eq:compat} holds and \(f\) has a single interior maximum, there are either one or two admissible roots in \((0,C/A)\); in that generic case the horizontal line \(y=\cos\theta_{12}\) intersects the two descending sides of \(f(x)\). In the perfectly symmetric case (\(r_1=r_2\)) the two roots are the daughter‑swap pair \(x\leftrightarrow y\) (see below).

\emph{Robust solver (bracketed).} Since \(\Phi\) does \emph{not} change sign at the interval endpoints, do \emph{not} attempt to bisect on \((0,C/A)\) directly. A stable procedure is:
\begin{enumerate}[itemsep=2pt,topsep=2pt,leftmargin=1.2em]
\item Sample \(f(x)\) on a fine grid over \((0,C/A)\); locate an interior maximizer \(x^\ast\) and record \(f_{\max}=f(x^\ast)\).
\item If \(f_{\max}<\cos\theta_{12}\), declare \emph{incompatibility} (no solution).
\item Otherwise, scan left from \(x^\ast\) until \(f(x)\) first drops below \(\cos\theta_{12}\); that defines a bracket \([x_L,x^\ast]\) containing the left root. Similarly, scan right to get \([x^\ast,x_R]\) for the right root (if present).
\item On each bracket where a sign change of \(\Phi\) is detected, use a monotone, bracketed 1D method (e.g., Brent’s) to obtain the root to tolerance.
\end{enumerate}
Under the single‑peak shape (unimodal \(f\)), this procedure recovers the 0, 1, or 2 admissible roots. If \(f\) is multimodal, the same steps return the roots adjacent to \(x^\ast\); additional roots (if present) can be found by repeating the search over disjoint sub‑intervals.

\paragraph*{Reconstruction.}
Given a root \(x^\star\), set
\[
y^\star = y(x^\star)=\frac{C-x^\star A}{B},\qquad
\frac{c_1}{c_0}=(x^\star)^2,\qquad
\frac{c_2}{c_0}=(y^\star)^2.
\]
Because only the \emph{ratios} are identified, the absolute scale of \((c_0,c_1,c_2)\) remains undetermined, as expected. If two roots are present, report both \(\bigl((x_1,y_1),(x_2,y_2)\bigr)\); they correspond to exchanging the daughter tariffs under the same observed geometry.

\paragraph*{Closed form in the near–symmetric case.}
If \(r_1=r_2\equiv r\) (hence \(A=B=r^{\alpha}\), \(W_1=W_2=r^{m}\)), set
\[
S\equiv \frac{C}{A}=\Bigl(\frac{r_0}{r}\Bigr)^{\!\alpha},\qquad
R\equiv \Bigl(\frac{r_0}{r}\Bigr)^{\!2m}.
\]
Then \(x+y=S\) from \eqref{eq:tom_wM}. Let \(p\equiv xy\). Using \(x^4+y^4=(x^2+y^2)^2-2x^2y^2=(S^2-2p)^2-2p^2\), equation \eqref{eq:tom_cos} reduces to a quadratic for \(p\):
\begin{equation}
\label{eq:p_quadratic}
(1+\cos\theta_{12})\,p^2\;-\;2S^2 p\;+\;\tfrac12\,(S^4-R)\;=\;0.
\end{equation}
Choose the physically admissible root \(p\in(0,S^2/4]\) (ensures \(x,y>0\)), then recover
\begin{equation}
\label{eq:x_y_from_p}
x,y=\frac{S\pm\sqrt{S^2-4p}}{2},
\qquad
\frac{c_1}{c_0}=x^2,\quad \frac{c_2}{c_0}=y^2.
\end{equation}
This symmetric closed form is numerically stable and makes explicit the daughter‑swap degeneracy \(x\leftrightarrow y\).

\paragraph*{Practical notes (conditioning and scaling).}
To improve numerical conditioning, it is beneficial to rescale by \(r_2\): set \(\tilde r_i=r_i/r_2\) so that \(B=1\) and \(W_2=1\). The admissible bracket then simplifies to \(x\in(0,C/A)\) with \(y=C-xA\). Evaluating \(\Phi\) via \eqref{eq:phi_def} in this normalized frame keeps all intermediate quantities \(\mathcal{O}(1)\) even when \(r_i\) span several decades. Measurement errors in \((r_0,r_1,r_2,\theta_{12})\) propagate to \((c_1/c_0,c_2/c_0)\) through \((A,B,C,W_1,W_2)\) and the scalar map \(x\mapsto\Phi(x)\); first–order sensitivities follow by differentiating \eqref{eq:phi_def} implicitly and using \eqref{eq:y_of_x}. When two solutions exist, the pair is typically close in near‑symmetric geometries; if an application requires a unique choice, an external prior (e.g., monotone relation between tariff and radius class) should be used to disambiguate.

\section{TUR remark}
\label{app:TUR_remark}
For steady Markovian dynamics, the thermodynamic uncertainty relation (TUR) bounds relative current fluctuations by the \emph{total} entropy production,
\[
\frac{\mathrm{Var}[J]}{\langle J\rangle^2}\;\ge\;\frac{2}{\Sigma_{\rm tot}/k_B}\,,
\]
see \cite{BaratoSeifert2015PRL,Gingrich2016PRL,HorowitzGingrich2020NatPhys}. Because \(\Sigma_{\rm tot}\ge \Sigma_{\rm info}\ge \zeta\,k_B\ln 2\int \dot{\mathcal I}\,dt\), any expression depending only on \(\int \dot{\mathcal I}\,dt\) and \(\zeta\) can at best \emph{upper-bound} this lower bound; i.e.\ \(2/(\zeta \ln 2\int \dot{\mathcal I}\,dt)\) is not a guaranteed constraint on fluctuations \emph{without further dynamical assumptions} (choice of current, Markovity/coarse-graining, etc.). For this reason, TUR is not used here to provide quantitative bounds written solely in the audit variables \(\int \dot{\mathcal I}\,dt\) and \(\zeta\).

\section*{Data availability}
This work uses the publicly available \textbf{High--Resolution Fundus (HRF)} retinal image corpus. The HRF Image Database is provided by the Pattern Recognition Lab, Friedrich--Alexander--Universität Erlangen--Nürnberg (FAU), under the Creative Commons Attribution 4.0 International (CC BY 4.0) license, which permits sharing and adaptation with appropriate credit and a link to the license~\cite{HRF_web}. When HRF is used to evaluate methods, Budai \emph{et al.} (IJoBI 2013) should be cited~\cite{Budai2013IJBI}; when segmentation labels are used, Odstr{\v c}il{\' i}k \emph{et al.} (IET Image Processing 2013) should be cited~\cite{Odstrcilik2013IETIP}. The dataset can be obtained from the HRF project page~\cite{HRF_web}. The present study does not redistribute HRF images.

\section*{Code availability}
All analysis code, configuration files, and figure–generation scripts used in this work are available in the public \emph{EPICproject} repository \cite{EPICprojectRepo}:
\href{https://github.com/JustinBen0131/EPICproject}{\texttt{github.com/JustinBen0131/EPICproject}}
(tag \texttt{v3-arxiv}). The repository contains the exact pipeline used to produce the included figures together with preregistration fragments and run–time parameters. A minimal node-level CSV bundle (radii, angles, per–node \(m\) and \(R(m)\), and QC flags) is provided as ancillary material to this submission.

\begin{acknowledgments}
The High--Resolution Fundus (HRF) images were created at the Pattern Recognition Lab, Friedrich--Alexander--Universität Erlangen--Nürnberg, with cooperating clinical partners acknowledged on the HRF project page. The author gratefully acknowledges these institutions and project maintainers for making the HRF data publicly available to the research community.
\end{acknowledgments}

\bibliography{ref}
\end{document}